\documentclass[cernpreprint, atlasdraft=false, texlive=2023, USenglish, texmf, orcidlogo]{atlasdoc}
 
\usepackage[subcaption,multirow=true]{atlaspackage}
\usepackage{atlasbiblatex}
 
\usepackage{atlasphysics}

\graphicspath{{logos/}{figures/}}
 
\usepackage{ANA-EXOT-2018-37-PAPER-defs}

\AtlasJournalRef{Phys. Rev. D 109 (2024) 112008}
\AtlasDOI{10.1103/PhysRevD.109.112008}

\AtlasTitle{Search for high-mass resonances in final states with a $\tau$-lepton and missing
transverse momentum with the ATLAS detector}

\AtlasAbstract{
 
A search for high-mass resonances decaying into a $\tau$-lepton and a neutrino using proton–proton collisions
at a center-of-mass energy of $\sqrt{s}=13$~TeV is presented. The full Run~2 data sample corresponding to an
integrated luminosity of $139~\text{fb}^{-1}$ recorded by the ATLAS experiment in the years 2015--2018 is
analyzed. The $\tau$-lepton is reconstructed in its hadronic decay modes and the total transverse momentum
carried out by neutrinos is inferred from the reconstructed missing transverse momentum.
The search for new physics is performed on the transverse mass between the $\tau$-lepton
and the missing transverse momentum. No excess of events above the
Standard Model expectation is observed and upper exclusion limits are set on the $W^\prime \to \tau\nu$ production
cross-section. Heavy $W^\prime$ vector bosons with masses up to $5.0$~TeV are excluded at $95\%$ confidence level,
assuming that they have the same couplings as the Standard Model $W$ boson. For non-universal
couplings, $W^\prime$ bosons are excluded for masses less than $3.5 \text{--} 5.0$~TeV, depending on the model
parameters. In addition, model-independent
limits on the visible cross-section times branching ratio are
determined as a function of the lower threshold on the transverse mass
of the $\tau$-lepton and missing transverse momentum.
}
 
\author{The ATLAS Collaboration}

\PreprintIdNumber{CERN-EP-2023-298}
 
\AtlasJournal{Physical Review D}

\hypersetup{pdftitle={ATLAS document},pdfauthor={The ATLAS Collaboration}}
 
\begin{document}
 
\maketitle
 
\tableofcontents

\section{Introduction}
 
Multiple theories beyond the Standard Model (SM) predict the existence of additional charged or neutral
heavy vector gauge bosons, \Wprime  or \Zprime,  that may be observable at the Large Hadron Collider (LHC).
The Sequential Standard Model (SSM) \cite{Altarelli1989} is a flavor-universal benchmark model that assumes the couplings
of the \Wprime and \Zprime bosons to fermions are identical to those of the $W$ and $Z$ bosons
in the SM. Other models that are investigated in this paper are referred to as non-universal gauge
interaction models (NUGIM)~\cite{sfm, TopFlavor,G221,Family221}, which can exhibit different couplings
for the three lepton generations.
 
Searches for new heavy gauge bosons decaying into $\tau$-lepton final states are mainly motivated by models that violate
lepton universality, such as the NUGIM. An  example is the extension of the electroweak gauge group by an additional
SU(2) symmetry group. The first and second generation of fermions can transform under one SU(2) symmetry group,
while the third generation of fermions can transform under the additional one.
At some energy scale, the extended symmetry is spontaneously broken to the SM electroweak
gauge symmetry and the \Wprime and \Zprime bosons appear as massive particles.
The non-universality of the \Wprime couplings to the SM fermions is parametrized by an angle parameter, \thetanu,
which is used to scale the couplings to the first and second generations of fermions by $\tan \thetanu$ and
the couplings to the third generation by $\cot \thetanu$. For $\cot \thetanu=1$, the model's couplings are identical to those of the SSM,
while values of $\cot \thetanu>1$ enhance the couplings to the third generation.
The total decay width relative to the \Wprime boson mass in the NUGIM depends on its mass and the value of $\cot\thetanu$,
and it spans from approximately 3\% in the SSM limit to 31\%--36\%, depending on the \Wprime mass, for larger $\cot\thetanu$.
 
Alternative models associate the two SU(2) groups with a left--right symmetry \cite{Chang:2018jko,Babu:2018vrl,Greljo:2018ogz}.
The left-symmetry exhibits itself in the electroweak theory and is associated with the parity violation observed in the weak
interaction. The right-symmetry has not been observed at the energies probed to date, thus it must be broken at some energy
scale and through spontaneous symmetry breaking the new, right-handed $\Wprime$ and $\Zprime$ bosons become massive.
 
The ATLAS experiment has searched for \Wprime bosons in the \textit{light-lepton}, $\Wprime \to \ell \nu$~$(\ell = e, \mu)$~\cite{EXOT-2018-30}, and $\tau$-lepton,
$\Wprime \to \tau \nu$~\cite{EXOT-2017-06}, channels.
The \textit{light-lepton} searches generally have a better sensitivity than $\Wprime \to \tau \nu$
for models with universal couplings to fermions because they suffer from less SM background and are enhanced by better
lepton reconstruction and identification efficiency. With data collected in the years 2015--2018, corresponding to an
integrated luminosity of $139$~fb$^{-1}$ recorded at a center-of-mass energy of $\sqrt{s} = 13$~\TeV, the ATLAS
\textit{light-lepton} search excludes \Wprime bosons in the SSM with masses up to $6.0$~\TeV~\cite{EXOT-2018-30} at 95\%
confidence level (CL).
The signatures of $\Wprime \to \tau \nu$ decays in the detectors at LHC are mainly the
hadronic decay products of a high-momentum $\tau$-lepton and large missing transverse
momentum, due to the prompt neutrino from the \Wprime decay and the neutrino from
the subsequent $\tau$-lepton decay.
The CMS experiment has searched for $\Wprime \to \tau \nu$ decays in data collected during the
years 2015--2018 and excluded \Wprime bosons in the SSM with masses up to \SI{4.8}{TeV}~\cite{CMS-2022ncp} at 95\% CL.
A similar search was performed by the ATLAS experiment in the data collected during the years 2015--2016, amounting to $36.1\text{ fb}^{-1}$ of
integrated luminosity, and excluded \Wprime bosons in the SSM with masses up to \SI{3.7}{TeV}~\cite{EXOT-2017-06} at 95\% CL.
 
Exclusion limits on the masses of the new heavy gauge bosons as a function of the parameter $\cot\thetanu$ can also be derived
by indirect searches. The most stringent limits are derived from electroweak precision measurements (EWPT)~\cite{G221LHC} and by the absence of lepton flavor
violation (LFV) in the SM~\cite{LeeLFV}. They exclude \Wprime bosons with masses below $1.8$--$2.5$~\TeV, depending
on the coupling. Weaker limits can be set either from tests of the unitarity of the Cabibbo--Kobayashi--Maskawa (CKM) matrix ~\cite{LeeCKM}
or $Z$-pole data~\cite{sfm}.
 
In this paper, a search for $\Wprime \to \tau \nu$ decays with the full Run 2 ATLAS data sample, corresponding to an integrated luminosity of $139$~fb$^{-1}$,
recorded with the ATLAS detector in proton--proton collisions at a center-of-mass energy of $\sqrt{s} = 13$~\TeV at the LHC is presented.
In this analysis, events were selected using missing transverse momentum triggers and
the $\tau$-leptons are exclusively reconstructed in their hadronic decay modes, which
account for 65\% of all $\tau$-lepton decays. Compared to the previous ATLAS search,
this analysis is making use of the additional data collected between the years 2017--2018,
the improvements in the reconstruction and identification of the $\tau$-leptons at ATLAS,
as well as the strategy of a multi-bin search approach.

\section{ATLAS detector}
 
\newcommand{\AtlasCoordFootnote}{
ATLAS uses a right-handed coordinate system with its origin at the nominal interaction point (IP)
in the center of the detector and the \(z\)-axis along the beam pipe.
The \(x\)-axis points from the IP to the center of the LHC ring,
and the \(y\)-axis points upwards.
Cylindrical coordinates \((r,\phi)\) are used in the transverse plane,
\(\phi\) being the azimuthal angle around the \(z\)-axis.
The pseudorapidity is defined in terms of the polar angle \(\theta\) as \(\eta = -\ln \tan(\theta/2)\).
Angular distance is measured in units of \(\Delta R \equiv \sqrt{(\Delta\eta)^{2} + (\Delta\phi)^{2}}\).}
 
The ATLAS detector~\cite{PERF-2007-01,ATLAS-TDR-19,PIX-2018-001} at the LHC is a multipurpose particle detector
with a forward--backward symmetric cylindrical geometry and a near \(4\pi\) coverage in
solid angle.\footnote{\AtlasCoordFootnote}
It consists of an inner tracking detector surrounded by a thin superconducting solenoid
providing a \SI{2}{\tesla} axial magnetic field, electromagnetic (EM) and hadron calorimeters, and a muon spectrometer.
The inner tracking detector covers the pseudorapidity range \(|\eta| < 2.5\).
It consists of silicon pixel, silicon microstrip, and transition radiation tracking detectors.
Lead/liquid-argon (LAr) sampling calorimeters provide EM energy measurements
with high granularity.
A steel/scintillator-tile hadron calorimeter covers the central pseudorapidity range (\(|\eta| < 1.7\)).
The endcap and forward regions are instrumented with LAr calorimeters
for both the EM and hadronic energy measurements up to \(|\eta| = 4.9\).
The muon spectrometer surrounds the calorimeters and is based on
three large superconducting air-core toroidal magnets with eight coils each.
The field integral of the toroids ranges between \num{2.0} and \SI{6.0}{\tesla\metre}
across most of the detector.
The muon spectrometer includes a system of precision tracking chambers and fast detectors for triggering.
A two-level trigger system is used to select events \cite{TRIG-2016-01,TRIG-2019-04}.
The first-level trigger is implemented in hardware and uses a subset of the detector information
to accept events at a rate below \SI{100}{\kHz}.
This is followed by a software-based high-level trigger (HLT) that
reduces the accepted event rate to \SI{1}{\kHz} on average
depending on the data-taking conditions.
 
An extensive software suite~\cite{ATL-SOFT-PUB-2021-001} is used in data simulation,
in the reconstruction and analysis of real and simulated data, in detector operations,
and in the trigger and data acquisition systems of the experiment.

\section{Signal and background samples}

Signal events of $\Wprime \to\tau\nu$ decays were generated at leading order with the \PYTHIA[8.212]~\cite{Sjostrand:2014zea}
event generator and the \NNPDF[2.3lo] parton distribution function (PDF) set ~\cite{Ball:2012cx_mine}. 
The A14 set of tuned parameters (\textit{tune})~\cite{ATL-PHYS-PUB-2014-021} was used to simulate the parton shower and
the hadronization process. The TAUOLA v2.9 package~\cite{TAUOLA} was used for the simulation of the $\tau$-lepton decays.
 
The \Wprime signal events were generated with invariant mass of the $\tau\nu$ system above \SI{25}{\GeV}. The event generation used an artificially
biased phase space sampling to generate more events at high invariant masses.
Signal events for various resonance masses in the range of \SI{500}{\GeV} to
\SI{6000}{\GeV} were modeled by reweighting this sample using a leading-order matrix-element (ME) calculation.
For the NUGIM signals, depending on the values of $\cot\thetanu$ and resonance mass, the total decay width
of the \Wprime bosons increased up to $36\%$, which
affects the signal acceptance. \Wprime decays into boson pairs ($WZ$, $Wh$) were
neglected in the calculation of the total decay width. This impacts the total decay width by less than $7\%$.
Very large values of $\cot\thetanu$ ($>5.5$) are not studied as the model becomes non-perturbative in this region.
The signal cross-section is corrected to account for next-to-next-to-leading-order (NNLO) quantum chromodynamics (QCD) effects by mass-dependent
$k$-factors. The NNLO QCD effects in the $k$-factors were calculated using VRAP v0.9~\cite{Anastasiou_2004} and the \CT[14nnlo] PDF set~\cite{Dulat:2015mca}.
Electroweak corrections as well as interference between $W$ and \Wprime bosons are expected to be model-dependent and are not considered.
 
The SM background to this search is divided into events where the selected $\tau$-lepton candidate originates from a quark- or gluon-initiated jet
({\em jet background}) and those where it does not ({\em non-jet background}). The jet background is primarily due to $W$/$Z$+jets, mainly \Znunujets,
and multijet production and is estimated from data. The non-jet background is
estimated by using simulation and mainly originates from $W \to \tau\nu$ production. Additionally, smaller contributions come from $W/Z/\gamma^*$ decays into leptonic final states, top-quark ($t\bar{t}$ and single top-quark) production, and diboson
($WW$, $WZ$, and $ZZ$) production, collectively referred to as {\em other background}. The generators and software packages used for
the simulation of the background samples are summarized in Table~\ref{tab:generators}. In addition, quantum electrodynamics (QED) final-state radiation in $W/Z/\gamma^*$ processes was simulated with \PHOTOSpp[3.52]
~\cite{Golonka:2005pn,Davidson:2010ew}. Decays of bottom and charm hadrons were simulated with \EVTGEN[1.2.0]
and \EVTGEN[1.6.0]~\cite{Lange:2001uf} for $W/Z/\gamma^*$ and top quark processes, respectively.
For diboson and \Znunujets production, virtual QCD corrections to matrix elements at next-to-leading-order (NLO) accuracy were provided by the
\OPENLOOPS library~\cite{Buccioni:2019sur,Cascioli:2011va,Denner:2016kdg}.
 
All simulated Monte Carlo (MC) events were processed through a simulation of the detector geometry and response~\cite{SOFT-2010-01}
using the \GEANTV{} framework~\cite{Agostinelli:2002hh}. The effect of multiple interactions in the same and neighbouring bunch
crossings (pileup) was modelled by overlaying the simulated hard-scattering event with
inelastic proton--proton ($pp$) events generated with \PYTHIA[8.186]~\cite{Sjostrand:2007gs}
and the A3 tune~\cite{ATL-PHYS-PUB-2016-017}. The average number of pileup
interactions (additional $pp$ collisions in the same or adjacent bunch crossings) observed in
the analyzed data is about $33$. All MC samples are reweighted so that the distribution of
the number of collisions per bunch crossing matches that observed in data. They are normalized using the integrated luminosity
of the data sample and the cross-sections calculated at NNLO in QCD and NLO for electroweak processes.
The NNLO QCD and NLO electroweak effects are included in the simulated $W/Z/\gamma^*$ samples by applying mass-dependent $k$-factors.
The electroweak correction effects are calculated using MCSANC~\cite{Arbuzov_2016} in the case of QED effects due to initial-state radiation,
interference between initial- and final-state radiation, and Sudakov logarithm single-loop corrections.
The software used for the reconstruction is the same for both simulated and
real data.

\begin{table}[h!]
\caption[Simulated samples.]{
Details of the generators and software packages used to simulate the background samples, including
the generation of the matrix element and the corresponding PDF set as well as
the modeling of non-perturbative effects such as parton showers, PDF set and MC tune.
}
\label{tab:generators}
\centering
\scriptsize
\begin{tabular}{llllll}
\toprule
Process    & Generator               &      ME order             & PDF            & Parton Shower               & Tune                                \\ \midrule
$W/Z$+jets & \POWHEGBOX[v1]~\cite{Nason:2004rx,Frixione:2007vw,Alioli:2010xd,Alioli:2008gx}     &        NLO                 & \CT[10nlo]~\cite{Lai:2010vv}       & \PYTHIA[8.186]                         &\AZNLO~\cite{STDM-2012-23}\\
&                                                                                    &                            &                                    & +\CTEQ[6L1]~\cite{Pumplin:2002vw}      &      \\
$t\bar{t}$ & \POWHEGBOX[v2]~\cite{Frixione:2007nw,Nason:2004rx,Frixione:2007vw,Alioli:2010xd}   &        NLO                 & \NNPDF[3.0nlo]~\cite{Ball:2014uwa_mine} & \PYTHIA[8.230]                         & A14  \\
&                                                                                    &                            &                                    &+\NNPDF[2.3lo]                          &      \\
Single top & \POWHEGBOX[v2]~\cite{Re:2010bp,Nason:2004rx,Frixione:2007vw,Alioli:2010xd}         &        NLO                 & \NNPDF[3.0nlo]                     & \PYTHIA[8.230]                         & A14  \\
&                                                                                    &                            &                                    &+\NNPDF[2.3lo]                          &      \\
 
Diboson    & \SHERPA[2.2.1] or 2.2.2~\cite{Bothmann:2019yzt}                                    &MEPS@NLO~\cite{Hoeche:2012yf}&\NNPDF[3.0nnlo]~\cite{Ball:2014uwa_mine}& \SHERPA ~\cite{Gleisberg:2008fv,Schumann:2007mg,Hoeche:2011fd,Hoeche:2012yf,Catani:2001cc,Hoeche:2009rj}                    & \SHERPA                  \\
\Znunujets& \SHERPA[2.2.1]          &       MEPS@NLO            & \NNPDF[3.0nnlo]& \SHERPA                     & \SHERPA                      \\
\bottomrule
\end{tabular}
\end{table}

\section{Analysis strategy and event selection}
\label{sec:analysis}

The efficient selection of high-mass $\Wprime\to\tau\nu$ decays and the suppression of backgrounds requires an accurate
identification of hadronic $\tau$-lepton decays, the reconstruction of their transverse
momentum and the reconstruction of the missing transverse momentum, which are used to
build the final discriminating variable for this search.
 
The interaction vertices from $pp$ collisions are reconstructed from inner detector
tracks with transverse momenta \pTtrk$>0.5$ \GeV that originate from the beam collision
region in the transverse plane. In presence of several primary vertices, the hard-scatter
primary vertex is chosen as the interaction vertex with the highest sum of squared \pTtrk.
Events with no hard-scatter primary vertex are rejected. Additionally, the candidate
events are required to satisfy standard data-quality criteria \cite{DAPR-2018-01}.

The $\tau$-lepton is reconstructed from its hadronic decay products.
Hadronic $\tau$-lepton decays are composed of a neutrino and a set of visible decay products (\tauhadvis),
typically one or three charged pions and up to two neutral pions.
The reconstruction of the visible decay products is seeded by jets reconstructed from topological clusters~\cite{PERF-2014-07} of energy
depositions in the calorimeter.
The reconstruction of \tauhadvis candidates is detailed in Refs.~\cite{PERF-2013-06,ATLAS-CONF-2017-029}.
 
The \tauhadvis candidates are required to have transverse momentum, $\pttauhadvis > \SI{30}{\GeV}$  and $|\eta|< 2.4$
(excluding the region $1.37 < |\eta| < 1.52$).
They are required to have one (1-prong) or three (3-prong) associated tracks within the {\em core region} of $\Delta R < 0.2 $ around the \tauhadvis
axis and an electric charge of $|Q|= 1$.  The prompt charged-particle tracks within the {\em core region} are classified using boosted decision trees (BDT)
against other tracks that can originate from other sources (such as tracks from photon conversions, underlying events or pileup),
which increases the reconstruction efficiency of the high-\pt \tauhadvis relative to the previous analysis.
 
Only the candidate with the highest transverse momentum in the event is selected.
 
Hadronic $\tau$-lepton decays are identified with a recurrent neural network (RNN) algorithm \cite{ATL-PHYS-PUB-2019-033} based on calorimetric
shower shape and tracking information to suppress backgrounds from quark- or gluon-initiated jets that are reconstructed as $\tau$-lepton candidates.
Since the jet background for high \Wprime masses is small, loose criteria are used for the $\tau$-lepton identification to maintain a high signal efficiency.
The chosen \textit{loose} working point of the $\tau$-lepton identification has an efficiency of $85\%$ for 1-prong candidates and $75\%$ for 3-prong candidates. At
the same time, the \textit{loose} working point provides a jet background rejection factor of at least 21 for 1-prong candidates and 90 for 3-prong candidates.
For the separation of 1-prong $\tau$-lepton candidates from misidentified electrons, a dedicated discriminant based on a BDT \cite{ATLAS-CONF-2017-029} is used with
an efficiency of $95\%$ and an electron background rejection factor of 50--100, depending on the pseudorapidity.
 
Events with a reconstructed electron~\cite{EGAM-2018-01} or muon~\cite{MUON-2018-03} are rejected.
This lepton veto makes use of \textit{loose} electron and muon identification criteria and considers
lepton candidates with transverse momenta above $20$~\GeV. The electrons are required to be reconstructed
with $|\eta|<2.47$ (excluding $1.37 < |\eta| < 1.52$) while muons are required to be reconstructed with
$|\eta|<2.5$. Electrons and muons are required to have tracks associated with the primary vertex with longitudinal impact
parameter $|z_0 \sin\theta|<$ \SI{0.5}{mm} and transverse impact parameter significance $|d_0/\sigma(d_0)|<5$ for
electrons and $|d_0/\sigma(d_0)|<3$ for muons, where $\sigma(d_0)$ is the measured uncertainty in $d_0$.
 
Jet candidates are reconstructed from topological clusters of energy deposition in the calorimeter~\cite{PERF-2014-07} using the anti-$k_t$
algorithm~\cite{Cacciari:2008gp,Fastjet} with a radius parameter $R = 0.4$.
They are calibrated using simulation with corrections obtained from in situ techniques
in data~\cite{JETM-2018-05}. Jets are required to have a transverse momentum above $20$~\GeV and
a pseudorapidity in the range of $|\eta| < 4.5$. To reduce the effect of pileup on jets with transverse momenta less than \SI{60}{GeV}, the jet vertex tagging
(JVT) and forward jet vertex tagging (fJVT) algorithms \cite{PERF-2014-03,PERF-2016-06} are used for jets with $|\eta| < 2.5$ and
$2.5<|\eta|<4.5$, respectively, to determine the likelihood of the jet originating from the hard-scattering vertex.
Jets enter indirectly in the analysis through the calculation of the missing transverse
momentum of the event.
 
The events are required to satisfy criteria designed to reduce non-collision backgrounds from cosmic rays, single-beam-induced events
and calorimeter noise~\cite{ATLAS-CONF-2015-029}. To further suppress single-beam-induced background, the \tauhadvis candidate
must have at least one associated track with transverse momentum greater than $10$~\GeV.
 
The missing transverse momentum with magnitude \met is calculated as the negative vectorial sum of the
transverse momenta of all reconstructed objects in the event. In addition,
a {\em soft term} \cite{ATLAS-CONF-2018-023} is added to account for the contribution from tracks that originate from the primary vertex
but are not associated with the reconstructed objects.
 
Events were selected by triggers that required \met to be above a threshold of $70$, $90$ or $110$~\GeV, depending on the data-taking period \cite{TRIG-2019-01}.
The offline-reconstructed \met is required to be at least $150$~\GeV, motivated by the high \Wprime masses ($\sim 0.5\text{--}6$~\TeV)
considered in this search, to suppress multijet background and to minimize the uncertainty in the trigger efficiency.
The trigger efficiency for this offline threshold is about $80\%$ and increases to more than $99\%$ for $\met > 200$~ \GeV.
Single $\tau$-lepton triggers were not used because they provide similar offline thresholds as the \met triggers, but also include a $\tau$-lepton
identification step at the HLT. The latter can result in a bias in the data selection and consequently the jet background estimation, which is
described in Section~\ref{sec:jetbackground}.
 
Correction factors are applied to simulation to account for differences observed in the detector response between data and MC.
This includes corrections for the $\tau$-lepton reconstruction and identification efficiencies, as well as the energy and momentum scales and resolutions of the
reconstructed objects~\cite{ATLAS-CONF-2017-029,JETM-2018-05,PERF-2016-07}. Additional corrections are derived for $\tau$-lepton candidates with $\pttauhadvis$ above \SI{100}{GeV}
that have a long flight length in the detector since the direct interaction of $\tau$-leptons with
the detector is not included in the simulation. More than 50\% of the $\tau$-leptons with $\pttauhadvis>$ $100$~\GeV\ decay after a flight distance of \SI{30}{mm}
in the transverse plane, which corresponds to the innermost layer of the ATLAS detector. The corrections are determined by comparing the reconstruction and identification
efficiencies for the selected $\Wprime\to\tau\nu$ sample with a dedicated simulated sample that includes direct interactions of $\tau$-leptons with the detector.
The missing transverse momentum trigger efficiencies are measured with \Zmumujets events, where the jet is reconstructed
as a $\tau$-lepton candidate, exploiting the fact that muons typically deposit only a small fraction of their energy in the calorimeter.
The trigger efficiencies are measured in data and simulation after removing the muons from
the \met calculation, which results in similar event topologies to those studied in this paper.
 
Additional kinematic selection criteria are applied to further reduce multijet background.
As the \Wprime bosons are typically produced with low transverse momenta, the $\tau$-lepton and the neutrino from the \Wprime decay
are produced back-to-back in the transverse plane and have approximately balanced transverse momenta.
Due to the large \Wprime mass and the resulting boost of the $\tau$-lepton and its decay products, the direction of the
neutrino from the $\tau$-lepton decay is close to the original direction of the $\tau$-lepton. The neutrino from the $\tau$-lepton decay
thus partially cancels the missing transverse momentum caused by the prompt neutrino from the \Wprime decay.
As a result, the \tauhadvis candidate and the missing transverse momentum are also preferentially balanced.
Therefore, the azimuthal angle between the $\tau$-lepton candidate and the missing transverse momentum, \deltaphi,
is required to be larger than $2.4$~radians and their transverse momentum ratio to be in the range of $0.7 < \pttauhadvis/\met < 1.3$.
 
The selection criteria discussed above define the {\em signal region} (SR) of the $\Wprime\to\tau\nu$ search.
The selection acceptance times efficiency, $\mathcal{A}\times\varepsilon$, is approximately 8\% for a low-mass resonance with \mwpssm $=$ \SI{500}{\GeV} and
is mainly affected by the trigger and \met requirements of the SR\@. For higher resonance masses near \SI{2.5}{\TeV},
the $\mathcal{A}\times\varepsilon$ increases to approximately 25\%. Due to the event selection requirements and the
increased production of off-shell \Wprime bosons at lower invariant mass, the selection $\mathcal{A}\times\varepsilon$ decreases
for \mwpssm above \SI{3}{\TeV} and reaches approximately 16\% for resonance masses at \SI{6}{\TeV}.
An overview of the selected event yields is given in Table~\ref{tab:selection}. A summary of the
event selection and the region requirements is given in Table~\ref{tab:selectionsummary}. In addition to the SR,
three {\em control regions} (CR1, CR2 and CR3), as defined in Section~\ref{sec:jetbackground}, are used to obtain a data-driven estimate of the
jet background and a {\em validation region} (VR) is used to validate the SM background in the high-\mt region.
 
The transverse mass of the reconstructed \tauhadvis candidate and \met, \mt, defined as
 
\begin{equation}
\mt = \sqrt{2 \met \pttauhadvis (1-\cos\deltaphi)}
\end{equation}

is used to further separate the signal from backgrounds as a high-mass \Wprime is expected to produce events at higher \mt values. Due to the presence
of the additional neutrino from the $\tau$-lepton decay, the distribution of the transverse mass for the signal process is not expected to show a Jacobian peak.
The separation between signal and background uses a profile likelihood based on the $\mt$ distribution, as described in Section ~\ref{sec:result}.

\begin{table}[]
\caption{Summary of the event selection requirements. The top part of the table summarizes the ``preselection'' requirements that apply to all regions used in this analysis. The bottom part shows the additional selection requirements for each individual region. Here, the symbol L stands for \textit{loose} $\tau$-lepton identification and  VL$\setminus$L denotes the requirement that the $\tau$-lepton candidate must satisfy the \textit{very loose} but fail to satisfy the \textit{loose} identification. }
\label{tab:selectionsummary}
\centering
\begin{tabular}{rccccc} \toprule
&  &    & \textbf{Preselection}  &   &  \\[0.1cm]\midrule
\met trigger &  &   & $70,90,110~$\GeV  &   &  \\[0.1cm]
Event cleaning &  &   & applied &   &  \\[0.1cm]
\tauhadvis tracks &  &   &     $1$ or $3$    &   &  \\[0.1cm]
\tauhadvis charge &  &   &      $\pm1$   &   &  \\[0.1cm]
$\pttauhadvis$ &  &   &      $>30$~\GeV   &   &  \\[0.1cm]
\tauhadvis $\pt^\text{leadTrack}$ &  &   & $>10$~\GeV &   &  \\[0.1cm]
Lepton veto &  &   & applied &   &  \\[0.1cm]
\deltaphi &  &   & $>2.4$~rad        &   &  \\[0.1cm]\midrule
&  &   &   \textbf{Region requirements}   &   &  \\[0.1cm]
& \multicolumn{1}{c}{\textbf{SR}}   & \multicolumn{1}{c}{\textbf{CR1}}  & \multicolumn{1}{c}{\textbf{CR2}}   & \multicolumn{1}{c}{\textbf{CR3}}  & \multicolumn{1}{c}{\textbf{VR}}  \\ \midrule
$\tau$-lepton identification & L  &   VL$\setminus$L  &     L    & VL$\setminus$L  & L \\[0.1cm]
\met & $>150$~\GeV & $>150$~\GeV & $<100$~\GeV & $<100$~\GeV & $>150$~\GeV \\[0.1cm]
$\pttauhadvis/\met$ & $\in [0.7,1.3]$  &   $\in [0.7,1.3]$  &   	$\cdots$      & 	$\cdots$  & $<0.7$ \\[0.1cm]
\mt & 	$\cdots$  &   	$\cdots$  &    	$\cdots$    & 	$\cdots$  & $>240$~\GeV \\[0.1cm]
\bottomrule
\end{tabular}
\end{table}

\begin{table}[h!]
\caption{Overview of the selected numbers of events for data, the SM backgrounds and a \WprimeSSM signal of mass $5$~\TeV.
The jet background is estimated from data and cannot be quantified before the requirements of $\met>150$~\GeV\ and
$\tau$-lepton identification. ``Preselection'' denotes all selection criteria described in Section~\ref{sec:analysis}
except for the $\tau$-lepton identification, \met and $\pttauhadvis/\met$ requirements. The last row summarizes the
number of observed and expected events above a large \mt threshold, but is not part of the SR selection.
The quoted uncertainties include both statistical and systematic sources of uncertainty.}
\label{tab:selection}
\centering
\small
 
\begin{tabular}{c|cccc|l}
\toprule
Selection                   &   Data    & $W\to\tau\nu$      & Jet background     & Other background     &   \WprimeSSM (\SI{5}{\TeV})\\ \midrule
Preselection                &
$3\,640\,749$ & $102\,000 \pm6\,000$               & 	$\cdots$                         &  $73\,000\phd\ph\pm6\,000\phd\ph  $   &  $18\phd\ph\pm5\phd\ph$       \\[0.1cm]
 
$\tau$-lepton identification&
$1\,189\,863$ &  $\ph84\,000 \pm 5\,000$           & 	$\cdots$                         &  $52\,000\phd\ph\pm4\,000\phd\ph$   &  $17\phd\ph\pm4\phd\ph$       \\[0.1cm]
 
$\met> 150$~\GeV            &
$\ph\phc\ph58\,528$   &  $\ph13\,400 \pm1\,600$    & $31\,000 \pm 9\,000$       &  $12\,000\phd\ph\pm1\,500\phd\ph  $   &  $15\phd\ph\pm4\phd\ph$       \\[0.1cm]
 
$0.7<\frac{ \pttauhadvis}{\met}<1.3$  &
$\ph\phc\ph18\,528$   &   $\ph\ph9\,700\pm1\,400$  & $\ph5\,800 \pm \ph\phc 400$&$\ph2\,900\phd\ph\pm\ph\phc500\phd\ph  $   &  $14\phd\ph\pm4\phd\ph$       \\[0.1cm]
 
$\mt>1$~\TeV	             &
$\ph\phc\ph\ph\ph\phc\ph58$&   $\ph\ph\phc\ph\ph 51\pm\ph\phc\ph12$    &  $\ph\ph\phc\ph 10\pm \ph\phc\ph\ph4$            &   $\ph\ph\phc\ph12.0\pm \ph\phc\ph\ph2.7$              &     $\ph7.2\pm3.3$             \\[0.1cm]
\bottomrule
\end{tabular}
\end{table}

\section{Jet background estimation}
\label{sec:jetbackground}
 
While all non-jet backgrounds in this analysis are estimated by using simulation, the jet background is estimated by using a data-driven
approach since the misidentification of jets as $\tau$-lepton candidates is not well modeled by the simulation.
In this approach, the jet background is determined from data events that fail to satisfy the $\tau$-lepton identification requirement of the signal region
and is transferred to the signal region using dedicated {\em transfer factors} measured in independent control regions.
 
Three control regions are defined.
The events in the first control region (CR1) are required to satisfy the same selection criteria as for the signal region, but
fail to satisfy \textit{loose} and satisfy \textit{very loose} $\tau$-lepton identification. The \textit{very loose} $\tau$-lepton identification working point
corresponds to a 95\% signal efficiency and 9.9 and 16 background rejection factors for 1-prong and 3-prong \tauhadvis candidates, respectively.
The other two control regions are enriched in dijet events.
For the definition of these regions, the requirement on $\pttauhadvis/\met$ is removed and the missing transverse momentum must not exceed $100$~\GeV,
while the other selection criteria remain the same as for the signal region. For one of these two control regions the $\tau$-lepton candidate must
satisfy \textit{loose} $\tau$-lepton identification (CR2), while for the other it must fail to satisfy \textit{loose} but satisfy \textit{very loose} $\tau$-lepton identification (CR3).
In all three control regions, the non-jet background is subtracted using simulation. The signal contamination in all three regions is small and is neglected.
 
\begin{figure}
\centering
\includegraphics[width=0.32\linewidth]{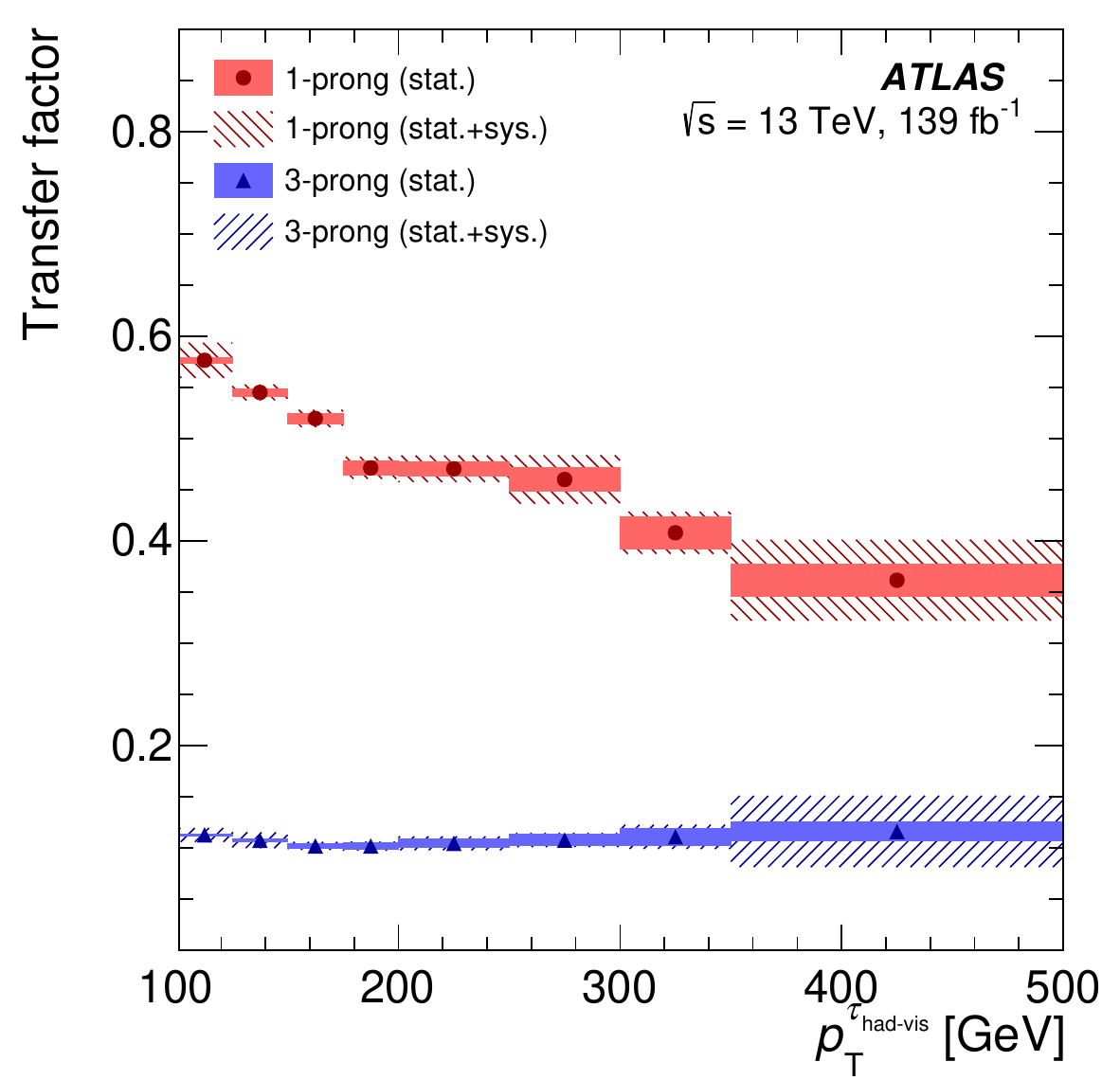}
\caption{Transfer factors used in the jet-background estimation as a function of $\pttauhadvis$, for 1-prong (circles) and 3-prong (upward triangles) $\tau$-lepton decays.
The uncertainty due to the limited number of events in CR2 and CR3 is shown as solid filled areas while the hatched areas indicates the
total statistical and systematic uncertainties.}
\label{fig:fakefactor}
\end{figure}

\begin{figure}
\centering
\begin{subfigure}{.32\linewidth}
\includegraphics[width=\linewidth]{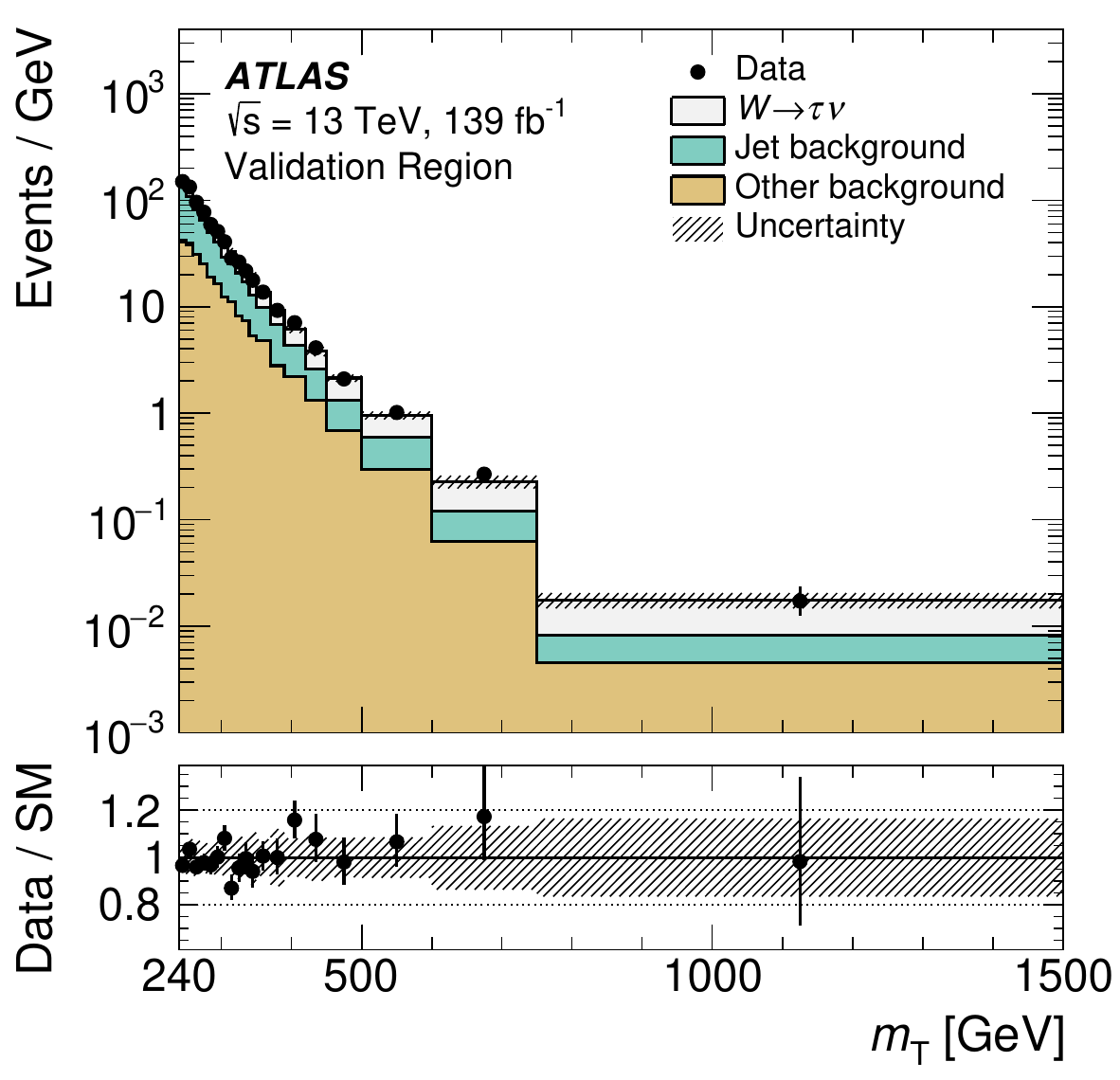}
\caption{\label{fig:SubVRKinematics_mt}}
\end{subfigure}
\begin{subfigure}{.32\linewidth}
\includegraphics[width=\linewidth]{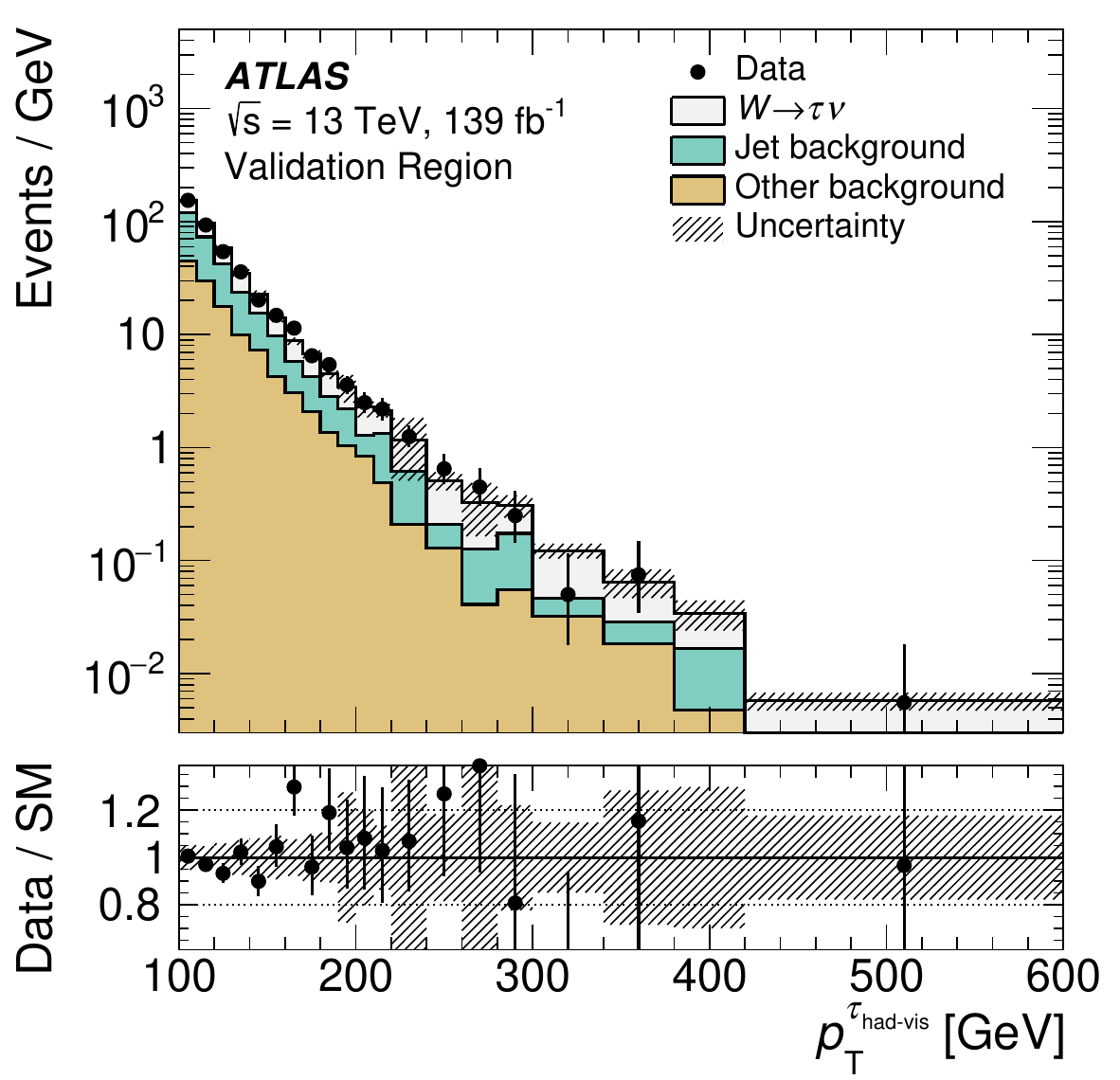}
\caption{\label{fig:SubVRKinematics_pt}}
\end{subfigure}
\begin{subfigure}{.32\linewidth}
\includegraphics[width=\linewidth]{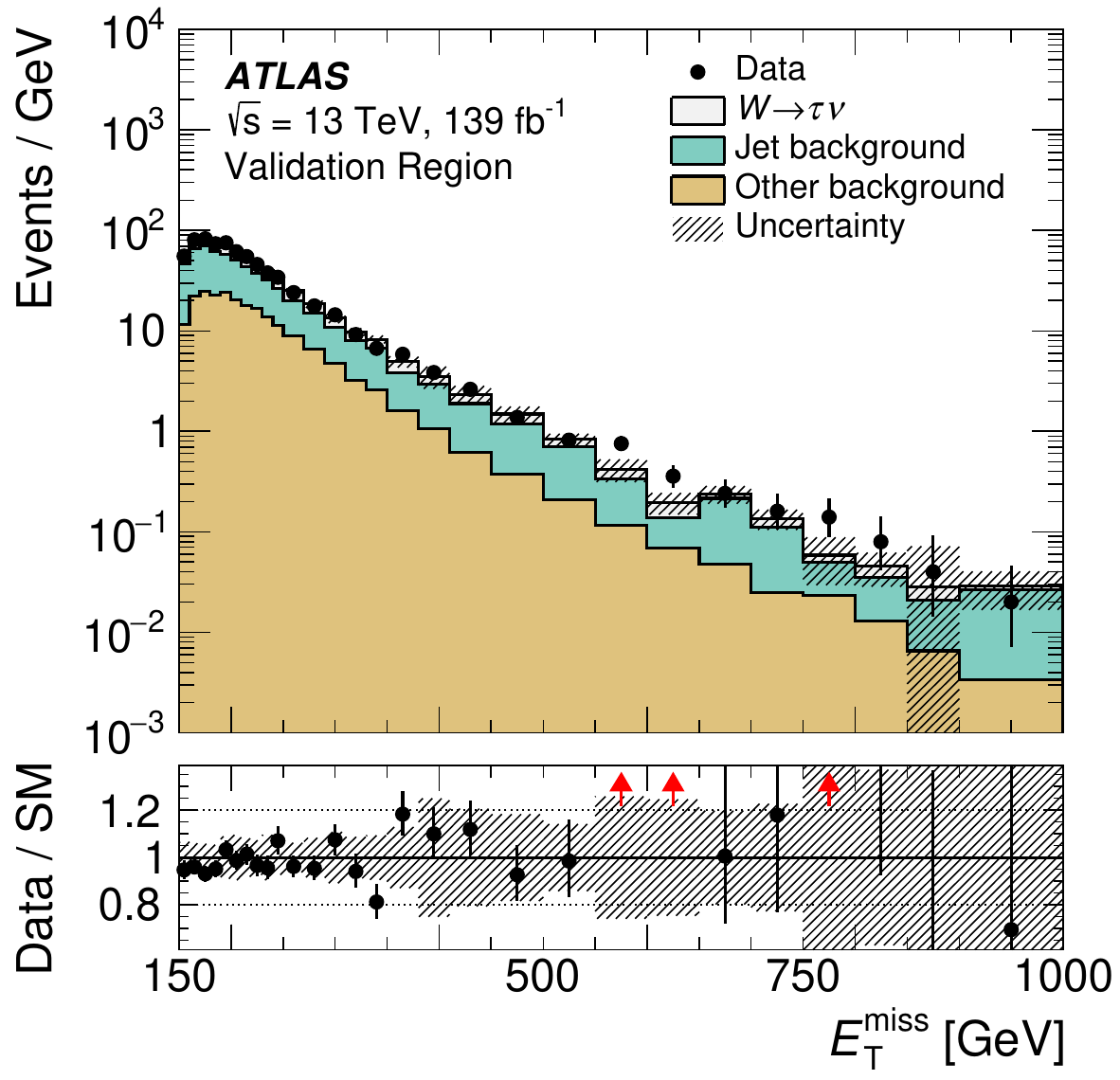}
\caption{\label{fig:SubVRKinematics_met}}
\end{subfigure}
\caption{Distributions of~\subref{fig:SubVRKinematics_mt} the transverse mass, \mt,~\subref{fig:SubVRKinematics_pt} the transverse momentum of the \tauhadvis candidate, $\pttauhadvis$, and~\subref{fig:SubVRKinematics_met} the missing transverse momentum in the event, \met,
in the validation region. The uncertainty band (hatched) shows the total statistical and systematic uncertainty. Arrows in lower panels indicate data points that lie outside the vertical range of the axis.}
\label{fig:VRKinematics}
\end{figure}

The transfer factors, $F_{ij}$, correspond to the ratio of events in CR2 and CR3 and are shown in Figure~\ref{fig:fakefactor}.
They are measured in intervals of $\pttauhadvis$ (denoted by index $i$), and separately for 1-prong and 3-prong $\tau$-lepton candidates (denoted by index $j$).
Further dependence on other observables, such as the \tauhadvis $\eta$ or the event's trigger, is found to give negligible effects and is not considered. The transfer factors
measured in the $\pttauhadvis$ interval of \SIrange{350}{500}{\GeV} are also used for reweighting events with $\pttauhadvis$ above \SI{500}{\GeV}.
The number of jet background events in the signal region, $N_\text{SR}^\text{jet}$, is computed from the number of data events in CR1
using these transfer factors:
 
\begin{equation}
N_\text{SR}^\text{jet} = \sum_{i,j} (N_{\text{CR1,}ij}^\text{data}-N_{\text{CR1,}ij}^\text{non-jet}) F_{ij} ~~\text{, where }
F_{ij}=\dfrac{ N_{\text{CR2,}ij}^\text{data}-N_{\text{CR2,}ij}^\text{non-jet} }{ N_{\text{CR3,}ij}^\text{data}-N_{\text{CR3,}ij}^\text{non-jet}}
\end{equation}
 
and $N_{X,ij}^\text{data}$ ($N_{X,ij}^\text{non-jet}$) corresponds to the number of data (simulated) events
populating the $i$-th and $j$-th intervals of \pttauhadvis and number of prongs, respectively, in the region
$X\in\{\text{CR1},\text{CR2},\text{CR3}\}$.
The jet background estimate is validated in a VR with
$\pttauhadvis/\met <0.7$ and $\mt>240$~\GeV. The data and estimated backgrounds are
found to be compatible within the uncertainties, as shown in Figure~\ref{fig:VRKinematics}.
 
\begin{figure}
\centering
\includegraphics[width=0.32\linewidth]{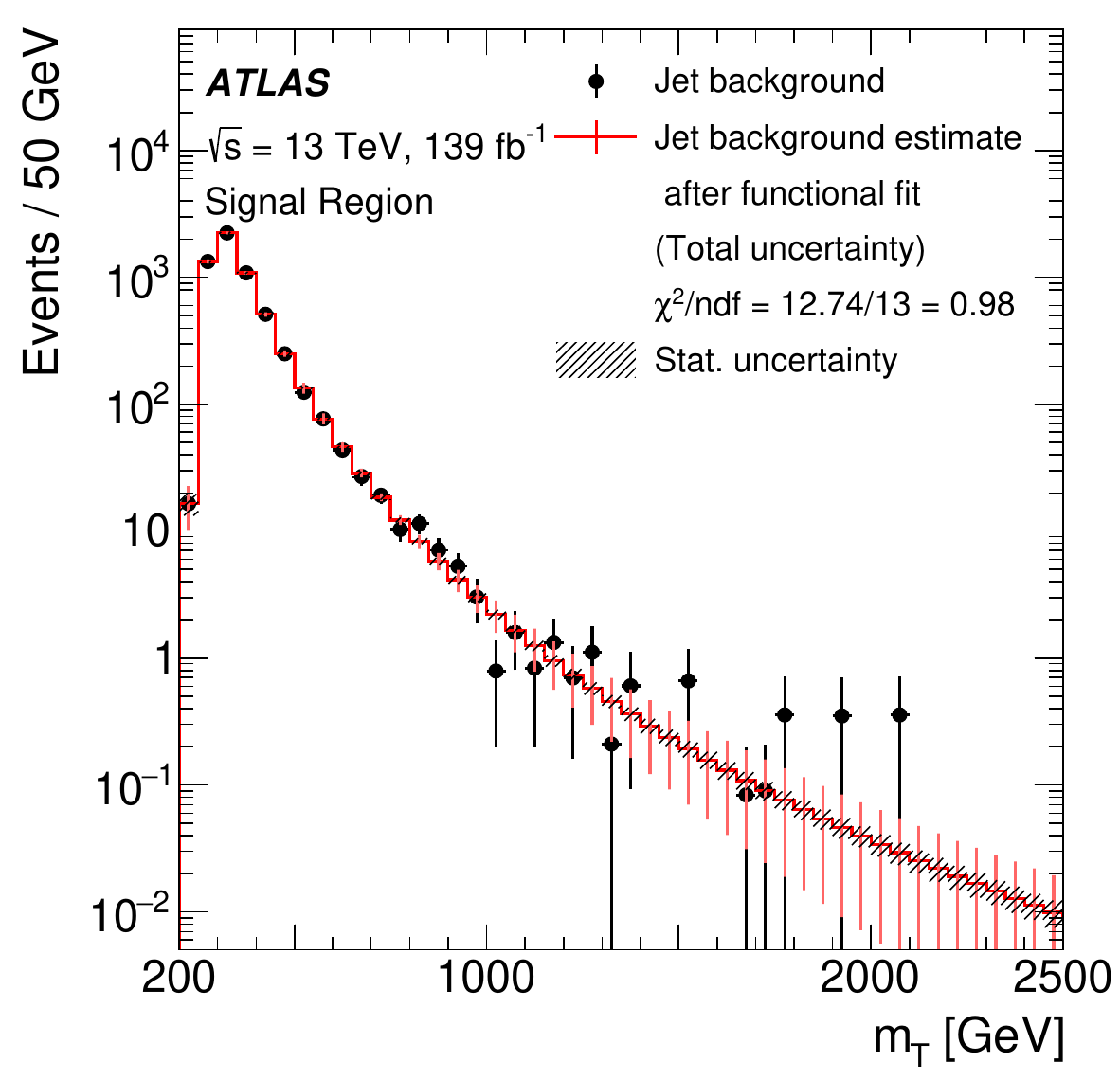}
\caption{Distribution of the transverse mass, \mt, of the jet background estimate.
The markers represent the jet background as estimated from applying the transfer
factor to the data in CR1. The line shows the binned jet background obtained
from the functional fit in the range $500$--$1200$ \GeV and its extrapolation for
$\mt$ values above $1200$~\GeV. The hashed area represents the statistical uncertainty and the vertical lines show
the total uncertainty of the jet background estimate.}
\label{fig:mtextrapolation}
\end{figure}

Due to the small number of events at high \mt in CR1, where the signal events are located,
a function of the form $f(\mt) = A \, \mt^B$ with free parameters $A$ and $B$
is fitted to the \mt distribution in the range of $450 < \mt < 1200$~\GeV.
This function is used  to parametrize the background shape for \mt$>500$~\GeV and is extrapolated for \mt$>1200$~\GeV.
Effects arising from kinematic suppressions, such as from the parton distribution functions, are neglected in this
simple extrapolation, as the jet background for \mt$>1200$~\GeV contributes to less than 15\% of the total SM background
and does not affect the analysis results.
Figure~\ref{fig:mtextrapolation} shows the \mt distribution of the jet background estimate before and after
performing the functional fit. The statistical uncertainty in the fitted function is determined by
pseudo-experiments from the jet background distribution and performing a fit with the same function.
 
\begin{figure}
\centering
 
\begin{subfigure}{.32\linewidth}
\includegraphics[width=\linewidth]{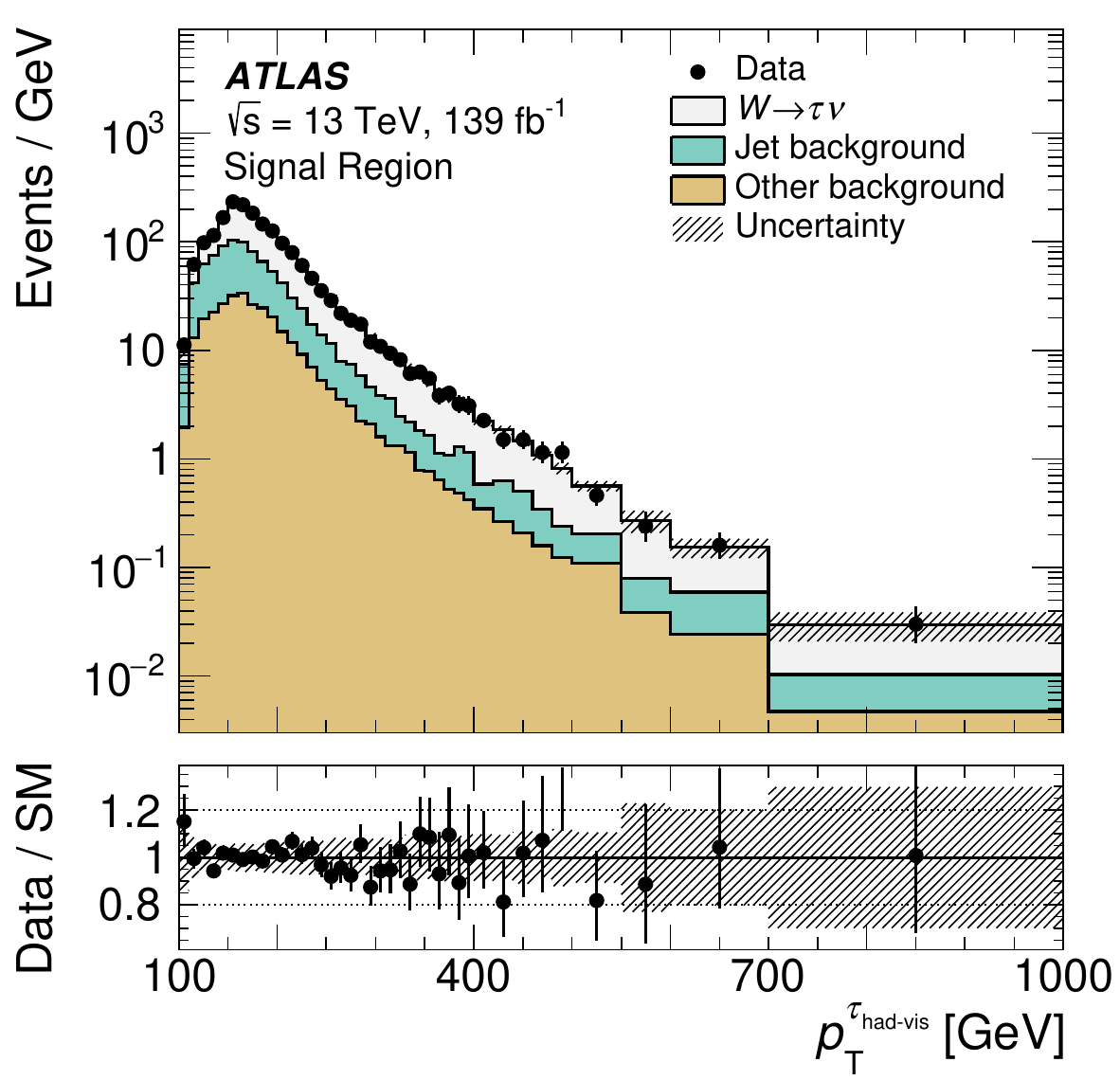}
\caption{\label{fig:SubKinematics_pt}}
\end{subfigure}
\begin{subfigure}{.32\linewidth}
\includegraphics[width=\linewidth]{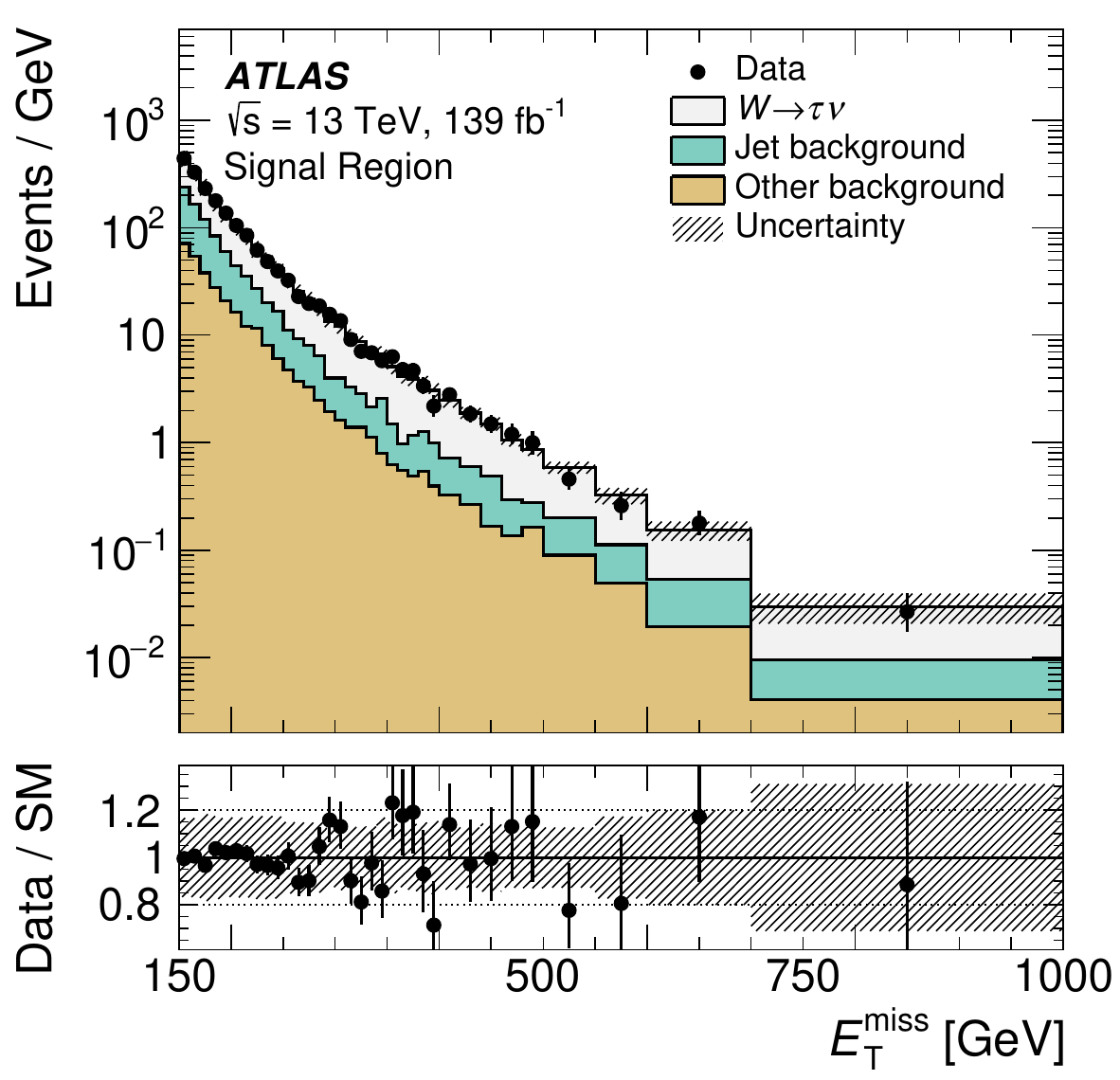}
\caption{\label{fig:SubKinematics_met}}
\end{subfigure}
\begin{subfigure}{.32\linewidth}
\includegraphics[width=\linewidth]{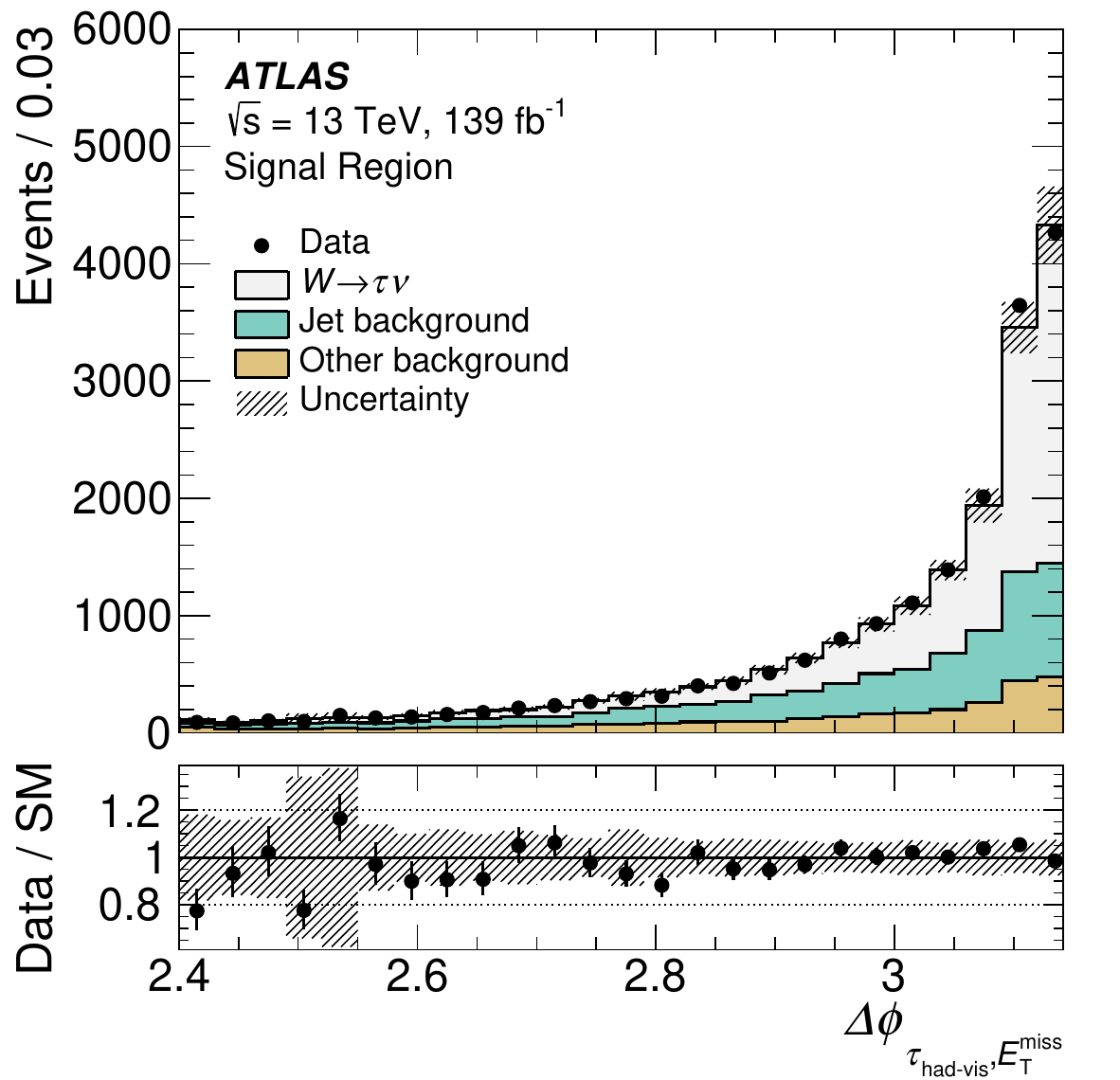}
\caption{\label{fig:SubKinematics_dphi}}
\end{subfigure}
 
\caption{Distributions of~\protect\subref{fig:SubKinematics_pt} the transverse momentum of the \tauhadvis candidate, $\pttauhadvis$,~\protect\subref{fig:SubKinematics_met} the missing transverse momentum in the event, \met,
and ~\protect\subref{fig:SubKinematics_dphi} the azimuthal angle between the \tauhadvis candidate and the missing transverse momentum, \deltaphi, in the signal region. The uncertainty band (hatched) shows the total statistical and systematic uncertainty.}
\label{fig:Kinematics}
\end{figure}

The data and estimated backgrounds are found to be compatible within the uncertainties for all observables used for the event selection.
Figure~\ref{fig:Kinematics} shows the distributions of the \tauhadvis transverse momentum,
the missing transverse momentum in the event, and the azimuthal angle between the \tauhadvis
candidate and the missing transverse momentum, in the signal region.
 
The estimation of the jet background is based on the assumption that the derived transfer factors calculated from control regions
are applicable for the signal region. Systematic uncertainties are assigned to account for any residual correlations between the transfer
factor and the \met criteria in the control region definitions that would arise if the jet composition differs between CR1 and CR2/CR3.
They are evaluated by repeating the jet background estimation using modified control region definitions.
The lower \met thresholds are varied from $0$ to $70$~\GeV and the upper threshold from $100$ to
$150$~\GeV, and the largest difference in each bin is taken as systematic uncertainty.
The statistical uncertainty in the transfer factor measurement is propagated from the control regions into the signal regions by using pseudo-experiments.
The uncertainty from the subtraction of non-jet contamination in the control regions is found to be small and corresponds to an uncertainty in the jet background yields
in the \mt distribution of $2\%\text{--}5\%$.
The uncertainty due to differences between the quark/gluon fraction of CR1 and CR3 is found to impact the jet background yields in the \mt distribution by $3\%\text{--}13\%$. It is determined by reweighting
the data in CR1 before obtaining the jet background estimate. The reweighting is performed such that the distribution of the $\tau$-lepton candidate's
jet seed width\footnote{The $\tau$-lepton candidate's jet seed width corresponds to the width of the jet that seeded the $\tau$-lepton reconstruction.}
has a similar shape for the two regions. The reweighting is also parametrized as a function of the $\tau$-lepton candidate's $\pttauhadvis$ and the number of associated
tracks used to capture the transfer factor differences.
 
The uncertainty from applying a constant transfer factor for $\pttauhadvis>350$~\GeV is determined from a recalculation of the transfer factors for an extension of the
measurement to transverse momenta of $1$~\TeV.
The uncertainty due to the extrapolation of the jet background estimate for $\mt > 500$~\GeV is evaluated by comparing the nominal estimate with
the one obtained when modifying the functional form using an alternative fit function ($f(\mt)=A \,\mt^{B+C\log \mt}$, with free parameters $A$,
$B$, and $C$), which also yields a good fit quality. Additional extrapolation uncertainties are considered by varying the lower and upper boundaries
of the fit range of the functional form by $\pm 50$~\GeV and rebinning the \mt distribution of the jet background before performing the functional fit.
The total uncertainty in the jet background estimate ranges from $4\%$ at $\mt = 200$~\GeV to $94\%$ at $\mt = 2$~\TeV, where, however, the jet background level is low.
For low \mt the uncertainty is mainly due to the subtraction of non-jet background contamination, while for high \mt, above \SI{800}{GeV}, is mainly due to the alternative fit function.
 
A summary of the systematic uncertainties in the jet background estimate for different \mt regions is given in Table~\ref{tab:jetsystematics}.

\begin{table}[h!]
\centering
\caption{ Summary of the uncertainties in the jet background estimate.
The ``$\cdots$'' symbol indicates that the uncertainty source is not applicable in
the  relevant \mt range.
}
\begin{tabular}{c|cc}
\toprule
\multirow{2}{*}{Systematic uncertainty} & \multicolumn{2}{c}{Relative uncertainty in the jet background [\%]}             \\[0.1cm]
& $\SI{200}{GeV} <\mt<\SI{300}{GeV}$  &          $\mt>\SI{2000}{GeV}$ \\[0.1cm]
\midrule
Non-jet background subtraction          &  $+$2/$-$3        & $+$2/$-$17 \\[0.1cm]
 
Variation of \met thresholds            &   $\pm$ 2       & $\pm$ 16 \\[0.1cm]
 
Quark/gluon ratio differences           &   $\pm$ 3       & $\pm$ 13 \\[0.1cm]
 
Extrapolation of transfer factor        &  $\pm$2         &  $\pm$16  \\[0.1cm]
 
Alternative fit function                &   	$\cdots$               &   $\pm$ 58     \\[0.1cm]
Lower fit range $\pm$\SI{50}{GeV}       &   	$\cdots$               &   $\pm$ 19     \\[0.1cm]
Higher fit range $\pm$\SI{50}{GeV}      &   	$\cdots$               &   $\pm$ 2      \\[0.1cm]
\mt rebinning                           &   	$\cdots$               &   $\pm$ 2     \\[0.1cm]
\bottomrule
\end{tabular}
\label{tab:jetsystematics}
\end{table}

\section{Systematic uncertainties}

The uncertainties in the data-driven estimate of the jet background have already been
discussed in Section \ref{sec:jetbackground}. In this section, the systematic
uncertainties due to reconstruction effects and the uncertainties of simulated
non-jet background modeling are discussed.
Specifically, uncertainties in the detector simulation impact the reconstruction,
identification and trigger efficiencies as well as the energy scales and resolutions
of reconstructed objects.
 
The uncertainty in the \tauhadvis energy scale is 3\%--4\%~\cite{ATLAS-CONF-2017-029}. The impact of this uncertainty is
found to vary with \mt, from $2\%$ at \mt$=200$~\GeV to $10\%$ at $\mt = 2$~\TeV\ for a signal with $\mwpssm=5$~\TeV, from $10\%$ to
$40\%$ for $W \to \tau \nu$ and from $15\%$ to $45\%$ for the other backgrounds. It is the overall largest systematic uncertainty
for the simulated backgrounds.
The uncertainty in the $\tau$-lepton identification efficiency is 5\%--6\%, as determined from measurements of  $Z \to \tau\tau$ events.
For higher transverse momenta outside of the range that can be probed with $Z \to \tau\tau$ decays, an additional uncertainty
that increases by $9\%$ per \TeV\ for 1-prong and $6\%$ per \TeV\ for 3-prong candidates is used, in accordance with studies
of high transverse-momentum jets~\cite{HIGG-2016-12}. The uncertainty in the electron veto efficiency is found to be $2\%$, independent of \mt.

The uncertainty in the factors used to correct for the absence of direct interactions of $\tau$-leptons with the detector material in
the simulation is due to the limited number of the generated events and to small differences between the nominal correction factors and
those obtained from an alternative simulated sample ($\gamma^* \to\tau\tau$).
This impacts the \mt distribution of the simulated signal and SM backgrounds,
with uncertainty that varies from $0.5\%$ at \mt$=200$~\GeV\ to $2\%$ at \mt$=3$~\TeV.
The \met trigger efficiency has an uncertainty of about $5\%$ for offline-reconstructed \met of $150$~\GeV\ that
decreases to below $1\%$ for larger \met values. It is determined by the statistical uncertainties in the trigger
correction factors measured in the \Zmumujets events and from the difference between correction factors measured in
\Wmunujets and $t\bar{t}$ events. The overall impact of the trigger scale factor uncertainties
(statistical and systematic) on the signal and background yields is $10\%$ at low \mt and becomes negligible above $1$~\TeV.
 
Although jets are not directly used in this analysis, their energy scale and resolution uncertainties affect the \met reconstruction
and lead to variations of the background yields of 1\%--2\% for signal and $W \to \tau \nu$ and 2\%--10\% for the other simulated backgrounds.
In addition, uncertainties related to the scale and resolution of the missing transverse energy soft term are evaluated.
They lead to variations of 0.5\%--2\% for signal and backgrounds, depending on \mt.
Uncertainties associated with the reconstruction of electrons and muons have a negligible impact.
 
The uncertainty of the combined integrated luminosity for the period 2015--2018 is 1.7\%~\cite{ATLAS-CONF-2019-021}, obtained
using the LUCID-2 detector~\cite{LUCID20} for the primary luminosity measurements. The impact of the uncertainties
on the pileup contribution in simulation yields a systematic uncertainty of less than $1\%$ for signal and simulated
backgrounds.
 
Theoretical uncertainties in the $W$ and $Z/\gamma^*$ differential production
cross-sections arise from PDF uncertainties, the uncertainty in the value of
the strong coupling constant, $\alphas$, and higher-order corrections. The strong
coupling constant is varied to $0.118$ \cite{PhysRevD.93.033006} from the nominal value of
$\alphas(m_Z) = 0.13$ used in the \CT[14nnlo] PDF set. Additional uncertainties
are estimated by simultaneously varying up and down the renormalization ($\mu_\text{R}$)
and factorization ($\mu_\text{F}$) scales of the \CT[14nnlo] PDF set by a
factor of two. A single nuisance parameter is used to quantify the uncertainty
due to the PDF. It is evaluated by the $90\%$ CL eigenvector variations of the
\CT[14nnlo] PDF as described in Ref.~\cite{Butterworth_2016}.
An additional uncertainty is derived due to the choice of the nominal PDF set,
by comparing the values of \CT[14nnlo] to those of ATLAS-epWZ16 \cite{STDM-2012-20}
and \NNPDF[3.0] PDFs, following standard prescriptions as in Ref.~\cite{Butterworth_2016}.
The maximum absolute deviation from the envelope of these comparisons is used as the
PDF choice uncertainty, when it is larger than the \CT[14nnlo] PDF eigenvector
variation envelope. The uncertainty in the electroweak correction is assessed by
comparing the multiplicative scheme ($(1+\delta_\text{QCD})\times(1+\delta_\text{EW})$)
with the additive scheme ($\delta_\text{QCD}+\delta_\text{EW}$), where the
NNLO QCD corrections, $\delta_\text{QCD}$, and the NLO electroweak corrections,
$\delta_\text{EW}$, are determined as described in Ref.~\cite{EXOT-2018-30}.
The additive approach is taken as the nominal value and its difference from
the multiplicative approach is taken as a symmetric uncertainty.
For the $Z/\gamma^*$ processes, an uncertainty to the photon-induced correction
is derived by the taking into account uncertainties of the photon PDF and quark
masses. The total cross-section uncertainty, once the theory uncertainties are
combined in quadrature, varies with the transverse mass from $5\%$ (at $\mt=200$~\GeV)
to $20\%$ (at $\mt=3$~\TeV) for $W \to \tau \nu$ and from $2\%$ to $20\%$ for the other
backgrounds, as they contain $W\to \ell \nu$ and neutral-current Drell--Yan processes.
For the $\Wprime\to\tau\nu$ signal, the theory uncertainty increases from $4\%$ to $50\%$
with $\mt$ and \Wprime mass. The theory uncertainty is not applied on the signal.
 
The $t\bar{t}$ cross-section has been calculated to NNLO and has an uncertainty of 5\%--6\%~\cite{Czakon:2011xx,Aliev:2010zk}.
The single-top cross-sections ($s$, $t$ and $Wt$ channels) have uncertainties of 3\%--5\%~\cite{Kant:2014oha,Kidonakis:2010ux}.
Uncertainties due to the modeling of hard scattering, fragmentation, interference and additional radiation for the top-quark processes
are also considered. They are determined by comparing alternative MC samples with different settings. A cross-section uncertainty
of $10\%$ is used for diboson production~\cite{Gleisberg:2008ta,Campbell:2011bn} and of $5\%$ for \Znunujets production~\cite{STDM-2016-01}.
All these result in a total uncertainty in the other background of approximately 10\%--15\%.
 
\begin{figure}
\centering
\includegraphics[width=0.6\linewidth]{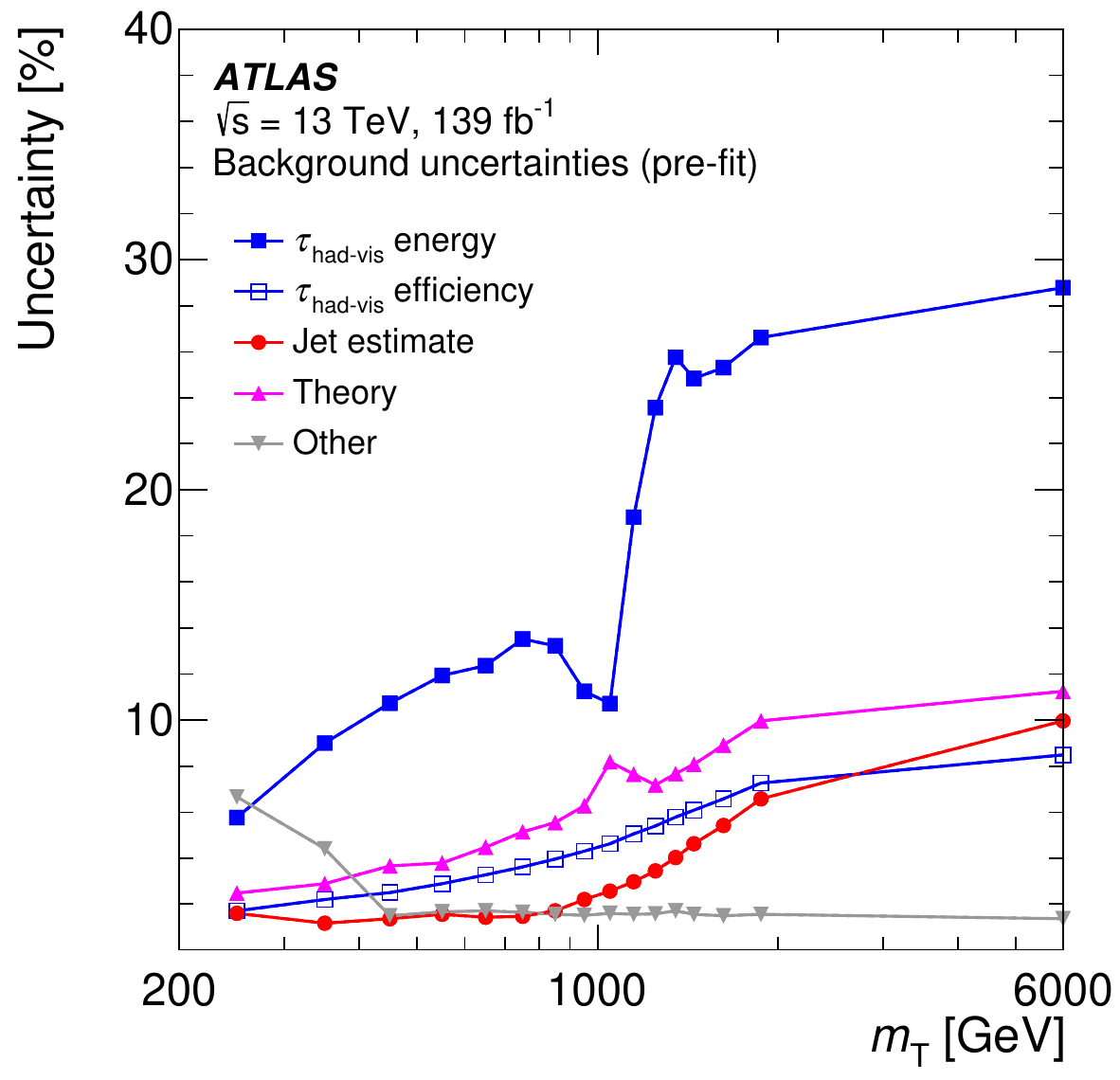}
\caption{ Relative systematic uncertainties (in \%) on the total background
yield before the likelihood fit to data (\textit{pre-fit}) as a function of \mt.
The filled squares show the uncertainty due to the $\tau$-lepton energy scale, the open squares show the
uncertainties originating from the $\tau$-lepton reconstruction and identification efficiencies, the
circles show the uncertainty originating from the jet background estimation, and the
upward triangles show the uncertainty from the simulated background cross-sections. Additional
sources of uncertainty, such as those from the trigger efficiency measurement, the jet energy scale and
the luminosity, are combined in the downward arrows.}
\label{fig:mTsystematics}
\end{figure}

The impact of these systematic uncertainties in the total background yields as a function of \mt is shown in Figure~\ref{fig:mTsystematics}.
 
\section{Results}
\label{sec:result}

For the statistical analysis of the data, a profile-likelihood fit to the \mt distributions of signal and background is performed.
For the signal, binned \mt distributions for a series of \Wprime masses in the range of $
\SI{500}{\GeV} < \mwp < \SI{6000}{\GeV}$ are used.
A profile-likelihood ratio is used as the test statistic. The likelihood functions in the ratio are products of Poisson probabilities
over all bins in the \mt distribution. Systematic uncertainties are included in the fit as nuisance parameters constrained by Gaussian prior probability density functions.
The theory uncertainties in the signal are not included as nuisance parameters, but are shown as an uncertainty band
on the predicted \Wprime cross-section.
In the numerator of the likelihood ratio, the likelihood function is maximized assuming the presence
of a signal above the expected background, while in the denominator the background-only hypothesis is assumed.
Upper $95\%$ CL limits on the signal normalization are derived using a modified frequentist confidence level
CL$_\text{s}=\displaystyle \frac{p_\text{s+b}}{1-p_\text{b}}~$~\cite{Read:2002hq}. The $p$-values for the signal-plus-background ($p_\text{s+b}$) and background-only ($p_\text{b}$)
hypotheses are determined using asymptotic formulae~\cite{Cowan:2010js}.
 
\begin{figure}
\centering
\includegraphics[width=0.6\linewidth]{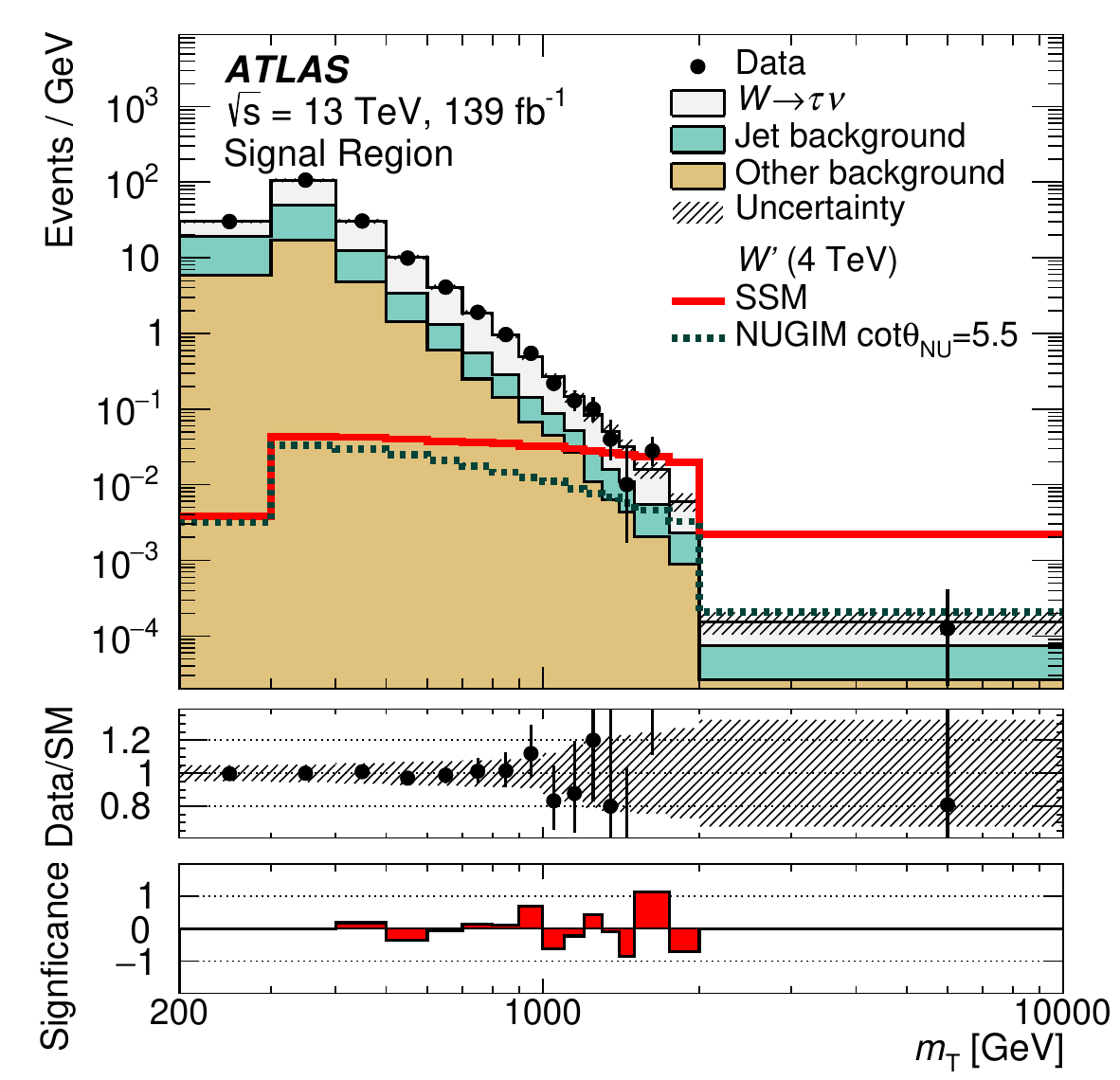}
\caption{Distribution of the transverse mass, \mt, in the signal region after the
likelihood fit to data (\textit{post-fit}) under the background-only hypothesis. The uncertainty band (hatched) shows the total statistical
and systematic uncertainty. The \mt distributions of a \Wprime signal with mass of \SI{4}{TeV} within the
SSM (red solid line) and NUGIM with $\cot\thetanu=5.5$ (dark green dotted line) are overlaid.
The significance of the data given the SM expectation and its uncertainty is given in the lower
panel. It is determined independently per bin and is computed as described in
Ref.~\cite{Diego:sig}.}
\label{fig:m_T}
\end{figure}

\begin{figure}
\centering
\begin{subfigure}[t]{0.4\linewidth}
 
\includegraphics[width=\linewidth]{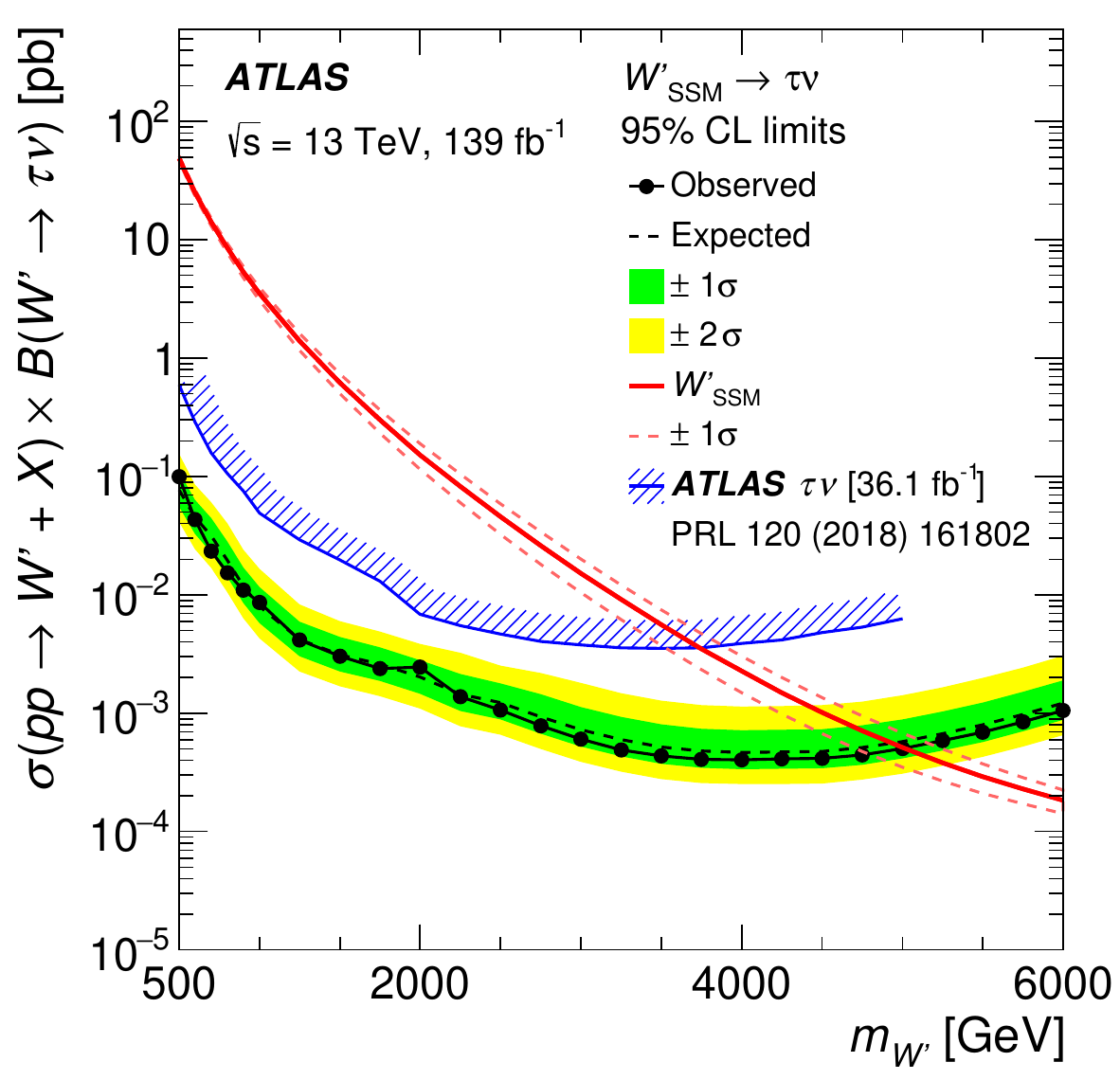}
\caption{\label{fig:Subfiglimit_SSM}}
\end{subfigure}
\begin{subfigure}[t]{0.4\linewidth}
\includegraphics[width=\linewidth]{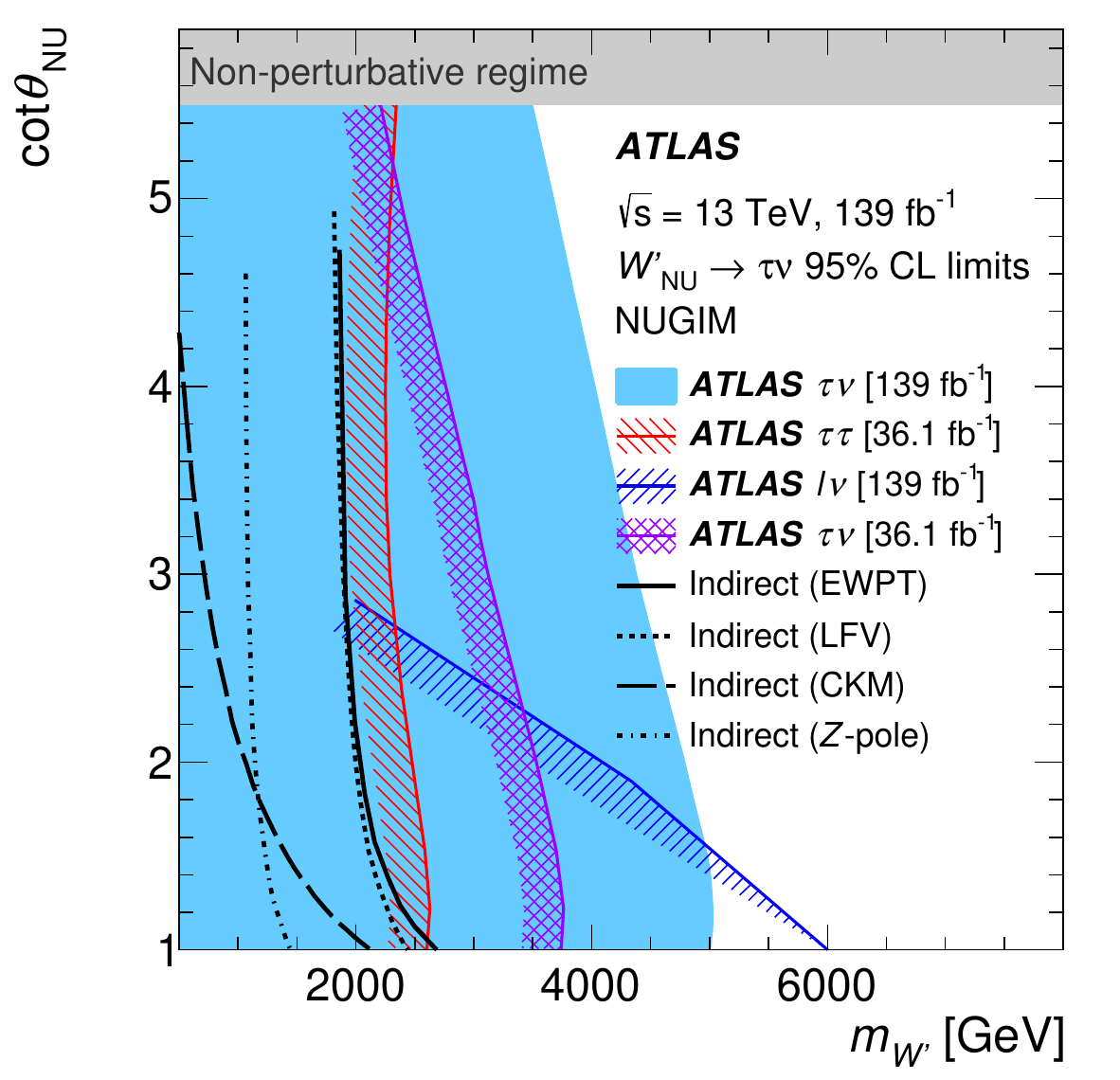}
\caption{ \label{fig:Subfiglimit_NUGIM}}
\end{subfigure}
 
\caption{~\subref{fig:Subfiglimit_SSM}
Observed (black markers) and expected (black dashed line) 95\% CL upper limits
on the cross-section times branching ratio ($\sigma \times \mathcal{B}$) as a
function of the \Wprime mass in the SSM.
The inner and outer bands show the $\pm 1$ and $\pm 2$ standard deviations,
respectively, of the expected limit.
The solid red line represents the theoretical cross-section and the dashed
red lines represent its theoretical uncertainty for the SSM signal.
The blue hatched line indicates the observed 95\% CL upper limits on
$\sigma \times \mathcal{B}$ of the previous ATLAS $\tau\nu$ \cite{EXOT-2017-06}
search.
~\subref{fig:Subfiglimit_NUGIM} Observed 95\% CL lower limit on the
\Wprime mass as a function of  the parameter $\cot \thetanu$ describing
the coupling to the third generation.
The blue shaded area represents the exclusion limits set by this analysis of the full
Run 2 data sample of ATLAS. For the same data sample, the exclusion limits set by the
$\Wprime\to\ell\nu$ search \cite{EXOT-2018-30} are also shown as blue forward hatched line.
The observed limits from the previous ATLAS $\tau\nu$ \cite{EXOT-2017-06}
(purple diagonal crosses) and $\tau\tau$~\cite{HIGG-2016-12} (red backward hatched
line) searches with $36.1~\text{fb}^{-1}$ are overlaid for comparison.
The \Wprime and \Zprime bosons are assumed to be degenerate in mass. Indirect limits at 95\% CL from fits to electroweak precision measurements (EWPT)~\cite{G221LHC},
lepton flavor violation (LFV)~\cite{LeeLFV}, CKM unitarity~\cite{LeeCKM} and the $Z$-pole data~\cite{sfm} are also overlaid.
}
 
\end{figure}

The \mt distribution after the profile-likelihood fit to data under the background-only hypothesis is shown in Figure~\ref{fig:m_T}.  There is good agreement between the data and total
background estimates. Since no significant deviation from the SM expectation is observed, upper limits on the cross-sections for the different
signal mass hypotheses are derived. The exclusion limits on the product of cross-section and branching fraction for $\Wprime\to\tau\nu$
as a function of the \Wprime mass in the SSM are shown in Figure~\ref{fig:Subfiglimit_SSM}. This search excludes the signal for \Wprime masses
up to $5.0$~\TeV\ at $95\%$ CL The expected limit is $4.9$~\TeV. For high signal masses ($>$\SI{4}{\TeV}), the validity of
the asymptotic formulae was tested against pseudo-experiments, resulting in an increase of the observed upper limits of less than 8\%.
The analysis for \Wprime boson masses above \SI{2}{\TeV} remains statistically limited and would benefit from increased integrated luminosity
and improved reconstruction of high-\pt \tauhadvis.
 
Upper exclusion limits on the production cross-section times branching ratio are also determined for NUGIM with $1 \le \cot\thetanu \le 5.5$.
\Wprime bosons with masses in the range of \SIrange{3.5}{5.0}{\TeV}, depending on $\cot\thetanu$, are excluded at $95\%$ CL as shown in Figure~\ref{fig:Subfiglimit_NUGIM}.
 
\begin{figure}
\centering
\includegraphics[width=0.4\linewidth]{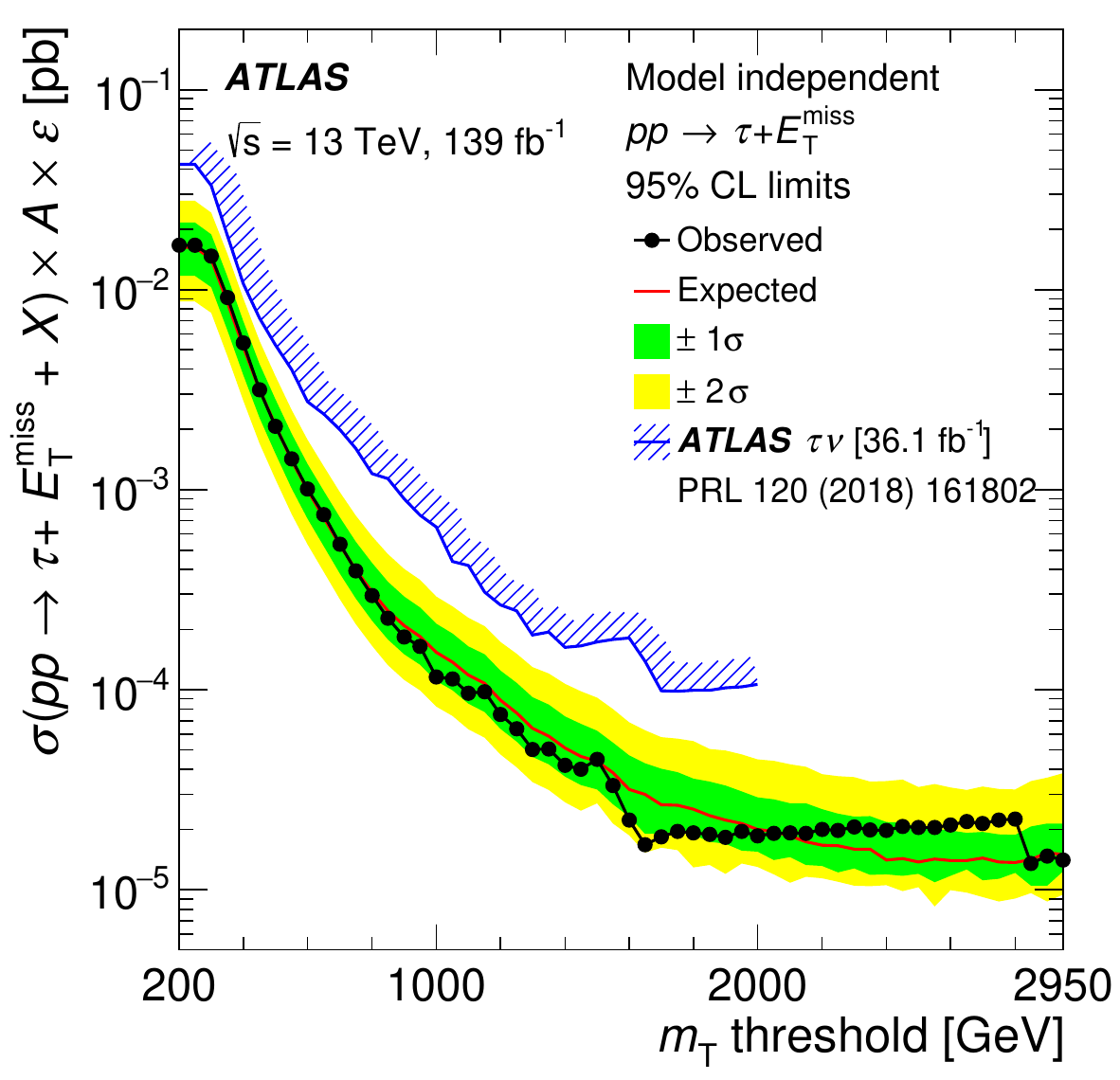}
 
\caption{Model-independent 95\% CL upper limits on the visible $\tau+\met$ cross-section as a function
of the transverse mass thresholds, \mtthres. The blue hatched line shows the
observed 95\% CL upper limits on the visible $\tau+\met$ cross-section of the previous ATLAS $\tau\nu$ \cite{EXOT-2017-06} search.
The steps in the observed upper limit at \mtthres of $1.6$ and $2.85$ TeV result of
the discrete nature of the data and that the highest \mt event
in data appeared with $\mt=2.83$ TeV.}
 
\label{fig:modelindependent}
\end{figure}

The calculation of the upper limits for the SSM and NUGIM assume certain shapes of the \mt distributions for the signal.
However, alternative models that also result in a $\tau$-lepton + \met final state can have different signal shapes.
For this reason, model-independent limits are provided, which are calculated as upper limits on the signal yields
above certain transverse mass thresholds, \mtthres, from $0.2$~\TeV\ to $2.95$~\TeV. For \mtthres above \SI{1.5}{\TeV}
the limits are calculated using pseudo-experiments because the expected and observed numbers of events drop considerably.
Figure~\ref{fig:modelindependent} shows the 95\% CL upper limits on the visible cross-section,
$\sigma(pp \to \tau+\met +X) \times \mathcal{A}\times\varepsilon$, as a function of \mtthres. Good agreement between the generated and reconstructed \mt distributions is observed
hence the \mtthres acceptance can be determined at generator level.
General models resulting in larger visible cross-sections for $\tau$-lepton + \met production than the determined upper limits are excluded by this analysis.
Thus, this analysis excludes possible signals with visible cross-sections larger than \SI{17}{fb} for \mtthres$=0.2$~\TeV
and \SI{0.014}{fb} for \mtthres$=2.95$~\TeV. Compared to previous results, this improves the upper limits on the visible cross-section for  \mtthres$>$\SI{1.5}{\TeV} by a factor of five.
 
The improvements in the limits compared to the previous analysis are mainly
due to the increased size of the data sample, an improved $\tau$-lepton
identification as well as the BDT-based track association to $\tau$-lepton
candidates and the multi-bin search approach.

\FloatBarrier

\section{Conclusion}
 
A search for $\Wprime \to \tau \nu$ decays in $139$~fb$^{-1}$ of $pp$ collisions at a center-of-mass energy of $\sqrt{s}=13$~\TeV
recorded with the ATLAS detector at the Large Hadron Collider is presented. The analysis is performed with hadronic $\tau$-lepton decays.
The signal is searched for in the transverse mass spectrum and no excess above the Standard Model expectation is observed.
Exclusion limits  are set on the cross-section for \Wprime production in the Sequential Standard Model as a function of the \Wprime mass.
\Wprime masses up to $5.0$~\TeV are excluded at 95\% confidence level.
This result improves upon the limits obtained in the previous $\Wprime\to\tau\nu$ analysis from ATLAS, based on a data sample of $36.1$~fb$^{-1}$,
by $1.3$~\TeV. \Wprime bosons in models with non-universal couplings are excluded for masses less than $3.5\text{--}5.0$~\TeV, depending on the model
parameters. Additionally, model-independent upper limits on the visible production cross-section for $\tau+\met$ are derived and range from \SI{17}{fb}
for a lower transverse mass threshold of $\mtthres = 200$~\GeV to \SI{0.014}{fb} for $\mtthres = 2.95$~\TeV.


We thank CERN for the very successful operation of the LHC and its injectors, as well as the support staff at
CERN and at our institutions worldwide without whom ATLAS could not be operated efficiently.
 
The crucial computing support from all WLCG partners is acknowledged gratefully, in particular from CERN, the ATLAS Tier-1 facilities at TRIUMF/SFU (Canada), NDGF (Denmark, Norway, Sweden), CC-IN2P3 (France), KIT/GridKA (Germany), INFN-CNAF (Italy), NL-T1 (Netherlands), PIC (Spain), RAL (UK) and BNL (USA), the Tier-2 facilities worldwide and large non-WLCG resource providers. Major contributors of computing resources are listed in Ref.~\cite{ATL-SOFT-PUB-2023-001}.
 
We gratefully acknowledge the support of ANPCyT, Argentina; YerPhI, Armenia; ARC, Australia; BMWFW and FWF, Austria; ANAS, Azerbaijan; CNPq and FAPESP, Brazil; NSERC, NRC and CFI, Canada; CERN; ANID, Chile; CAS, MOST and NSFC, China; Minciencias, Colombia; DNRF and DNSRC, Denmark; IN2P3-CNRS and CEA-DRF/IRFU, France; SRNSFG, Georgia; BMBF, HGF and MPG, Germany; RGC and Hong Kong SAR, China; ISF and Benoziyo Center, Israel; INFN, Italy; MEXT and JSPS, Japan; CNRST, Morocco; NWO, Netherlands; RCN, Norway; MEiN, Poland; FCT, Portugal; MNE/IFA, Romania; MESTD, Serbia; MSSR, Slovakia; ARRS and MIZ\v{S}, Slovenia; DSI/NRF, South Africa; MICINN, Spain; SRC and Wallenberg Foundation, Sweden; SERI, SNSF and Cantons of Bern and Geneva, Switzerland; MOST, Taipei; STFC, United Kingdom; DOE and NSF, United States of America.
 
Individual groups and members have received support from BCKDF, CANARIE, CRC and DRAC, Canada; PRIMUS 21/SCI/017 and UNCE SCI/013, Czech Republic; COST, ERC, ERDF, Horizon 2020 and Marie Sk{\l}odowska-Curie Actions, European Union; Investissements d'Avenir Labex, Investissements d'Avenir Idex and ANR, France; DFG and AvH Foundation, Germany; Herakleitos, Thales and Aristeia programmes co-financed by EU-ESF and the Greek NSRF, Greece; BSF-NSF and MINERVA, Israel; Norwegian Financial Mechanism 2014-2021, Norway; La Caixa Banking Foundation, CERCA Programme Generalitat de Catalunya and PROMETEO and GenT Programmes Generalitat Valenciana, Spain; G\"{o}ran Gustafssons Stiftelse, Sweden; The Royal Society and Leverhulme Trust, United Kingdom.
 
In addition, individual members wish to acknowledge support from CERN: CERN \& Society Foundation (ATLAS PhD Grant); Chile: Agencia Nacional de Investigaci\'on y Desarrollo (FONDECYT 1190886, FONDECYT 1210400); China: National Natural Science Foundation of China (NSFC-12075060); Czech Republic: PRIMUS Research Programme (PRIMUS/21/SCI/017); EU: H2020 European Research Council (H2020-MSCA-IF-2020: HPOFHIC - 10103); European Union: European Research Council (ERC - 948254), Horizon 2020 Framework Programme (MUCCA - CHIST-ERA-19-XAI-00); France: Agence Nationale de la Recherche (ANR-21-CE31-0022); Germany: Deutsche Forschungsgemeinschaft (DFG - CR 312/5-1); Italy: Istituto Nazionale di Fisica Nucleare (FELLINI G.A. n. 754496); Poland: Polish National Agency for Academic Exchange (PPN/PPO/2020/1/00002/U/00001), Polish National Science Centre (NCN UMO-2019/34/E/ST2/00393, UMO-2020/37/B/ST2/01043); South Africa: South Africa National Research Foundation (CPRR118515); Spain: Generalitat Valenciana (APOSTD/2019/165,PGC2018-094856-BI00, Artemisa, FEDER, IDIFEDER/2018/048), La Caixa Banking Foundation (La Caixa Foundation, LCF/BQ/PI19/11690014, LCF/BQ/PI20/11760025), Ministry of Science and Innovation (RYC2022-038164-I), PROMETEO and GenT Programmes Generalitat Valenciana (CIDEGENT/2019/023, CIDEGENT/2019/027, CIDEGENT/2019/029, GVA-SEJI/2020/037); Sweden: Swedish Research Council (SRC - 2017-05160, VR 2017-05092, VR 2018-00482), Knut and Alice Wallenberg Foundation (KAW 2017.0100, KAW 2018.0157, KAW 2018.0458, KAW 2019.0447); Switzerland: Swiss National Science Foundation (SNSF - PCEFP2\_194658); United Kingdom: Leverhulme Trust (Leverhulme Trust RPG-2020-004).


\clearpage

\printbibliography

\clearpage
 
 
\begin{flushleft}
\hypersetup{urlcolor=black}
{\Large The ATLAS Collaboration}

\bigskip

\AtlasOrcid[0000-0002-6665-4934]{G.~Aad}$^\textrm{\scriptsize 100}$,
\AtlasOrcid[0000-0002-5888-2734]{B.~Abbott}$^\textrm{\scriptsize 117}$,
\AtlasOrcid[0000-0002-7248-3203]{D.C.~Abbott}$^\textrm{\scriptsize 101}$,
\AtlasOrcid[0000-0002-2788-3822]{A.~Abed~Abud}$^\textrm{\scriptsize 34}$,
\AtlasOrcid[0000-0002-1002-1652]{K.~Abeling}$^\textrm{\scriptsize 53}$,
\AtlasOrcid[0000-0002-2987-4006]{D.K.~Abhayasinghe}$^\textrm{\scriptsize 92}$,
\AtlasOrcid[0000-0002-8496-9294]{S.H.~Abidi}$^\textrm{\scriptsize 27}$,
\AtlasOrcid[0000-0001-5329-6640]{H.~Abramowicz}$^\textrm{\scriptsize 148}$,
\AtlasOrcid[0000-0002-1599-2896]{H.~Abreu}$^\textrm{\scriptsize 147}$,
\AtlasOrcid[0000-0003-0403-3697]{Y.~Abulaiti}$^\textrm{\scriptsize 5}$
% ...
\bigskip
\\

% ...
$^{5}$High Energy Physics Division, Argonne National Laboratory, Argonne IL; United States of America.\\
% ...
$^{27}$Physics Department, Brookhaven National Laboratory, Upton NY; United States of America.\\
% ...
$^{34}$CERN, Geneva; Switzerland.\\
% ...
$^{53}$II. Physikalisches Institut, Georg-August-Universit\"{a}t G\"ottingen, G\"ottingen; Germany.\\
% ...
$^{92}$Department of Physics, Royal Holloway University of London, Egham; United Kingdom.\\
% ...
$^{100}$CPPM, Aix-Marseille Universit\'e, CNRS/IN2P3, Marseille; France.\\
$^{101}$Department of Physics, University of Massachusetts, Amherst MA; United States of America.\\
% ...
$^{117}$Homer L. Dodge Department of Physics and Astronomy, University of Oklahoma, Norman OK; United States of America.\\
% ...
$^{147}$Department of Physics, Technion, Israel Institute of Technology, Haifa; Israel.\\
$^{148}$Raymond and Beverly Sackler School of Physics and Astronomy, Tel Aviv University, Tel Aviv; Israel.\\
% ...

\end{flushleft}


\end{document}